\documentclass[review, 11pt]{elsarticle}
\usepackage{lineno,hyperref}
\usepackage{setspace}
\modulolinenumbers[5]
\journal{European Journal of Operational Research}
\usepackage{float}
\usepackage{geometry}
\usepackage[utf8]{inputenc}
\usepackage{indentfirst}
\usepackage{graphicx}
\usepackage{makecell}

\usepackage{algorithm}
\usepackage{algorithmic}
\usepackage{amsmath}
\usepackage{amsfonts}
\usepackage{booktabs}
\usepackage{multirow}
\usepackage{threeparttable}
\usepackage[figuresright]{rotating}

\usepackage{tikz}
\usetikzlibrary{arrows,graphs}
\usetikzlibrary{chains, decorations.pathreplacing, positioning} 
\usetikzlibrary{positioning} 
\usetikzlibrary{shapes.geometric}
\usepackage{pgf}
\usetikzlibrary{arrows,automata}
\usepackage{verbatim}
\usepackage[position=top]{subfig}
\usepackage{natbib}
\usepackage{amsmath, amsthm}

\usepackage{smartdiagram}
\usepackage{url}

\usepackage{titlesec}
\titleformat{\chapter}
{\normalfont\Large\bfseries}{Appendix \thechapter}{1em}{}
\titlespacing*{\chapter}{0pt}{-1.5em}{2em}

\begin{document}
	
	\setstretch{1.175}

	\begin{frontmatter}
		
		\title{The Dial-a-Ride Problem with Limited Pickups per Trip}
		
		\cortext[cor1]{Corresponding author: Roel Leus (\texttt{roel.leus@kuleuven.be})}
		
		\author[ku]{Boshuai Zhao}
		\author[thu]{Kai Wang}
		\author[bjtu]{Wenchao Wei}
		\author[ku]{Roel Leus \corref{cor1}}
		
		\address[ku]{Research Center for Operations Research \& Statistics, KU Leuven, Belgium}
		\address[thu]{School of Vehicle and Mobility, Tsinghua University, China}
		\address[bjtu]{School of Economics and Management, Beijing Jiaotong University, China}

		\begin{abstract}
			
			The Dial-a-Ride Problem (DARP) is an optimization problem that involves determining optimal routes and schedules for several vehicles to pick up and deliver items at minimum cost. Motivated by real-world carpooling and crowdshipping scenarios, we introduce an additional constraint imposing a maximum number on the number of pickups per trip. This results in the Dial-a-Ride Problem with Limited Pickups per Trip (DARP-LPT)\@. We apply a fragment-based method for DARP-LPT, where a fragment is a partial path. Specifically, we extend two formulations from \citet{Rist2021}: the Fragment Flow Formulation (FFF) and \textcolor{black}{the Pickup-Space Fragment Formulation (PSFF)\@. Furthermore, our results show that PSFF outperforms FFF, which in turn surpasses traditional arc-based formulations in both solution quality and computational efficiency. Additionally, we compare several existing fragment sets that differ in the length of their partial paths and find that the sets with shorter partial paths yield the best solution times when used with PSFF\@. In addition, we propose a new mixed fragment set, which is useful when the sets with longer partial paths become too large. In such cases, it yields the lowest CPU time.}

		\end{abstract}
		
		
	\end{frontmatter}

	\section{Background}\label{background}
	
	The Dial-a-Ride Problem (DARP) is an optimization problem that involves finding optimal routes and schedules for a fleet of vehicles to pick up and deliver a set of items (passengers or parcels). Each item has a specific pickup and delivery location and time window, and the objective is to schedule the vehicles in a way that minimizes the total travel time or distance, while satisfying various operational constraints such as vehicle capacity, time windows, and pickup and delivery locations.
	
	DARP has found extensive applications in the sharing mobility \citep{Mourad2019,Tafreshian2020} and food delivery sectors. Within the sharing mobility domain, the ride-sharing routing problem represents a variant of DARP\@. This problem focuses on optimizing routes for drivers, allowing them to efficiently reach their destinations while accommodating passenger pickups and drop-offs, all within the minimum possible time or cost \citep{Dessouky2022}.
	Similarly, in the context of online food delivery, the central system needs to make decisions on how to serve submitted requests by efficiently picking up and delivering orders with the goal of minimizing delivery time, which can be seen as an online variant of DARP \citep{GUO202213}.
	
	A frequently disregarded yet vital aspect in this type of routing problems is the limit on the number of pickups per trip. Many studies tend to focus on constraints such as time windows and maximum ride duration, always assuming these can substitute for pickup limits. Nevertheless, explicitly incorporating restrictions on the number of pickups can be crucial for optimizing delivery operations, especially within the shared mobility sector.
	For instance, in the context of carpooling, each driver typically accommodates a maximum of four passengers per trip, considering vehicle seat capacity and passenger convenience. Service providers such as 
	UberPool\footnote{\url{https://www.uberguide.net/how-many-people-can-ride-in-an-uber/}}, 
	DiDi\footnote{\url{https://web.didiglobal.com/au/help-center/how-many-people-are-allowed-on-a-didi-share-ride/}}, 
	and BlaBlaCar\footnote{\url{https://support.blablacar.com/hc/en-gb/articles/360014470360}} 
	enforce such restrictions, allowing drivers to pick up no more than two to four passengers on a single trip. Indeed, optimizing a route by initially picking up one passenger and dynamically delivering other passengers, ultimately reaching the first passenger's destination, is technically feasible. However, private mobility companies often avoid this strategy due to considerations related to passenger comfort.
	Another influential factor in this consideration is driver preference. In food delivery platforms like Meituan\footnote{\url{https://finance.sina.com.cn/tech/2022-05-10/doc-imcwipii9054944.shtml}}, 
	drivers possess the autonomy to determine the highest number of orders they undertake at one time. 
	Similarly, in the crowdshipping domain, considering drivers' preferences is crucial as they have the right to set their preferred maximum number for parcel pickups \citep{Arslan2019}.
	In particular, a company should not anticipate an occasional driver, who might only intend to deliver a few parcels on a sporadic basis or is simply looking to earn some extra income, to handle a high volume of parcels like a full-time staff driver. A similar assumption is found in the concept of limited delivery stops as discussed in \citet{Simoni2020} and \citet{Fessler2022}. In this paper, a ``customer'' can entail more than one person and/or item, but gives rise to only one pickup and two stops (one for pickup and one for delivery).

	In summary, although most previous research has ignored explicit limitations on the number of pickups per trip, in reality such limits have a significant impact on improving customer satisfaction and accommodating driver's preference.
	To address such a carpooling scenario where a fleet of vehicles needs to pick up and deliver a group of passengers with the goal to minimize cost, and each vehicle has its unique origin and destination and adheres to a maximum customer pickup limit, this paper introduces an innovative problem termed the Multi-depot Dial-a-Ride Problem with Limited Pickups per Trip (MDARP-LPT)\@. 	
	We also introduce a variant called the Dial-a-Ride Problem with Limited Pickups per Trip (DARP-LPT), where all vehicles start and end their routes at the same depot. DARP-LPT is applicable in crowdshipping scenarios, where several occasional riders are used to handle pickup and delivery orders, such as food or parcel deliveries. These riders (e.g.\ students or part-time riders) do not have specific origins and destinations for travel; their goal is to complete a limited number of orders as a part-time job for additional income. Unlike carpooling, food and parcel deliveries typically occur within a specific physical area, which is why the problem assumes the use of a central depot.
	
	Since MDARP-LPT and DARP-LPT share the same problem structure, we collectively refer to them as (M)DARP-LPT\@. \textcolor{black}{To efficiently address (M)DARP-LPT, we propose a unified approach featuring two distinct fragment-based formulations and evaluate different types of fragment sets as input. Both formulations are adapted from the DARP framework of \citet{Rist2021}. The first converts fragment-based decision variables into a fragment–vehicle–based form, a necessary adjustment even for (single-depot) DARP-LPT\@. The second introduces a novel pickup-space fragment formulation, extending the network with a pickup count dimension, similar to time in the time-space formulation, and which uses pickup-space decision variables. As for fragment sets: traditional methodologies employ several distinct sets, each with fragments (partial paths) of different levels of length. We also introduce a novel mixed fragment set, combining fragments of different lengths to reduce the total number while preserving useful ones as much as possible.}
	
	From a practical perspective, this research addresses a commonly overlooked aspect in the dial-a-ride problem: the limited number of pickups per trip. Our developed method offers more tailored and efficient solutions for the problem, thereby contributing to the improvement of carpooling and crowdshipping services. From an academic perspective, \textcolor{black}{this study proposes novel extended fragment-based formulations and conducts a comprehensive comparison and evaluation of multiple fragment sets for routing problems. Our findings offer valuable insights for addressing various problems that incorporate pickup and delivery. Additionally, this study is the first to explore the combination of different fragment types for constructing the fragment set used in the formulations, providing valuable insights for routing problems.}
	
	The remainder of this article is organized as follows: Section~\ref{liter} provides a comprehensive literature review. Following that, Section~\ref{DARP-LPT} formally defines (M)DARP-LPT, presenting an arc-based formulation\textcolor{black}{, two fragment-based formulations, and a path-based formulation}. Section~\ref{fragmethod} introduces traditional and novel methods for fragment generation. Then, in Section~\ref{numericaldiscussion}, we analyze the results of our computational experiments. Finally, we draw conclusions in Section~\ref{conclusion}.

	\section{Literature review}	\label{liter}
	
	Since (M)DARP-LPT is a variant of DARP, we investigate existing research methodologies on related optimization problems, such as the Pickup and Delivery Problem (PDP) and the Dial-a-Ride Problem. PDP involves transporting a set of items from their pickup locations to their respective delivery locations (where multiple items from a single pickup node may be destined for different delivery nodes), with the objective of finding optimal routes while minimizing transportation costs \citep{doi:10.1287/trsc.29.1.17}. DARP is a special case of PDP where each location corresponds to a single item, and every item is associated with a unique pickup and delivery location pair.

	Traditional methods, like branch and cut \citep{doi:10.1287/opre.1060.0283} and branch and price \citep{doi:10.1287/trsc.1090.0272,doi:10.1287/opre.1100.0881}, have been utilized for PDP and DARP, but recent studies suggest that fragment-based methods exhibit better performance. Here, a fragment is a partial path where only the start and end node have an empty load.
	
	A notable example of the fragment-based approach is the work of \citet{Alyasiry2019}, who apply a fragment-based strategy to address the Pickup and Delivery Problem with Time Windows, following the Last-In-First-Out loading strategy. \textcolor{black}{Examples can be represented as $p1p2p3d3d2d1$ or $p1p2d2p3d3d1$, where ``$p$'' denotes a pickup node, ``$d$'' denotes a delivery node, and the number indicates the customer index. This representation will also be used throughout the remainder of the paper. By combining fragments, \citet{Alyasiry2019} are able to enumerate all feasible paths; based on this, they construct a fragment-based formulation on a time-space network.} The fragments' applicability extends to various related problems, retaining the characteristic of empty-loaded start and end nodes but not limited to the specific \textcolor{black}{Last-In-First-Out setting. Subsequently, these fragments are referred to as ``full fragments''. In this context,} \citet{Su2023} effectively compute the minimum excess user ride time in an electric autonomous DARP using the fragment-based method.

	Building on \citet{Alyasiry2019}, \citet{Rist2021} propose the use of restricted fragments to address the dial-a-ride problem. \textcolor{black}{Since the full fragments in this problem don't necessarily adhere to the Last-In-First-Out property, their number would be quite high. However, \citeauthor{Rist2021} find that a full fragment with, for instance, the form ``$ppdpdpdd$'' can be decomposed into the restricted fragments ``$ppd$'' +``$pd$''+ ``$pdd$'', which start at a pickup node whose predecessor is a delivery node (or depot) and ends at a delivery node whose successor is a pickup node (or depot). Note that ``$ppdpdpdd$'' is merely an illustration where the last pickup does not necessarily precede the first delivery, this does not imply a universal pattern. Rather, it is intended to contrast with the form ``$ppppdddd$'', where the last pickup must precede the first delivery.} Obtaining these restricted fragments involves initially deriving them from fragments conforming to the form ``$ppppdddd$'' and then applying a dominance rule to minimize their count. \textcolor{black}{Unlike \citet{Alyasiry2019}, \citet{Rist2021} formulate the model on a physical network instead of a time-space network. Time-related constraints are handled purely via callbacks, yet the computational burden remains modest.}
	
	\textcolor{black}{Subsequently, \citet{Rist2022} introduced the concept of extended full fragments for the Selective Dial-a-Ride Problem (SDARP), a variant of DARP that aims to maximize passenger service. An extended full fragment is defined as the concatenation of a full fragment followed by an empty arc. Notably, the instances used in SDARP feature stricter time windows and shorter ride times than those in standard DARP\@. These tighter constraints motivate the use of a time-space network and help ensure a manageable number of extended full fragments. \citeauthor{Rist2022}'s \citeyearpar{Rist2022} computational results demonstrate that extended full fragments outperform existing fragment types for such tightly constrained instances. Similarly, \citet{Zhang2022} adopt a similar definition of extended full fragments, where each fragment is formed by concatenating an empty arc followed by a full fragment. They  tackle a vehicle-customer coordination problem that shares similarities with DARP but incorporates the flexibility for passengers to move in order to meet their drivers faster. In line with \citet{Rist2022}, the test instances in \citet{Zhang2022} feature even tighter time windows and ride time constraints (e.g., 1.2 to 1.5 times the delivery time limit of an ideal detour-free trip), which further contributes to keeping the number of fragments manageable. To tackle this problem, they design an algorithm based on a time-space fragment-based network. Their procedure not only achieves effective results, but also demonstrates its value for online DARP.}
	
	While recent literature has made significant contributions to fragment-based methods, \textcolor{black}{none can be directly applied to handle the limited number of pickups in our problem statement. 
		Additionally, different fragment types may result in varying computational performance. T}o the best of our knowledge, no existing study has systematically compared different fragment types or explored the potential of combining multiple fragment categories.
	
	\textcolor{black}{Another recent development is the event-based formulation for DARP \citep{Gaul2022, Gaul2024}, where an event represents a vehicle's state  as defined by its location and onboard passengers. The event-based formulation is quite similar to the traditional arc-based formulation of \citet{doi:10.1287/opre.1060.0283}, but in the event-based model the graph is constructed with events as nodes and feasible transitions as arcs. This design implicitly enforces capacity, precedence, and pairing constraints, thereby reducing the number of explicit constraints and improving computational efficiency compared to traditional arc-based models. Nevertheless, events inherently provide less routing information than fragments, as noted by \citet{Gaul2024}. Given the fairly superior performance of the fragment-based method on the benchmark instances of \cite{Rist2021}, we only concentrate on fragment-based strategies in this study.}

	\section{Problem statement and linear formulations}\label{DARP-LPT}

	Both MDARP-LPT and DARP-LPT are extensions of DARP, and their major difference from DARP lies in the limitations on the number of pickups per trip. In this section, we begin by providing a formal problem description of DARP-LPT and a\textcolor{black}{n} NP-hardness proof in Section~\ref{problemDef}. Subsequently, we present several linear formulations for DARP-LPT: an arc-based formulation in Section~\ref{ABF}, \textcolor{black}{two fragment-based formulations in Section~\ref{fragmentformulate}, and a path-based formulation in Section~\ref{PBF}.} The arc-based formulation closely resembles the one outlined in \citet{doi:10.1287/opre.1060.0283}, except for the constraints on the limited number of pickups. \textcolor{black}{The two fragment-based formulations build upon the framework of \citet{Rist2021}. The first formulation extends the framework by transitioning from fragment-based decision variables to fragment–vehicle-based ones. The second extends the network into a pickup–space representation, introducing pickup–space fragment-based variables to account for pickup constraints. This adaptation is tailored to our specific requirements for limited number of pickups. The path-based formulation presented in Section~\ref{PBF} adopts a set partitioning structure, as in \citet{doi:10.1287/trsc.1090.0272}.} Finally, in Section~\ref{MDARPProbForm}, we introduce MDARP-LPT, illustrating the differences in its definition and formulations as compared to DARP-LPT.

	\subsection{Problem description}\label{problemDef}
	
	DARP-LPT involves $n$ customers, with each customer $i$ having a unique pickup and delivery location pair represented as $(i, i+n)$, where $n$ is the total number of customers. The pickup and delivery locations can be represented by the sets $P=\{1,2,\dots,n\}$ and $D=\{n+1,n+2,\dots,2n\}$, respectively. The location set $N=P\cup D \cup \{0, 2n+1\}$ consists of all pickup and delivery locations, along with the origin depot location $0$ and the destination depot location $2n+1$. Each location~$i\in P\cup D$ has a specific time window $(e_i, l_i)$, defining the earliest and latest time a vehicle can arrive at that location.
	The arc set is denoted by $A$, including all arcs  $(i,j)\in N \times N$ with $i\ne j$, and the cost and time associated with traversing the arc $(i, j)\in A$ are represented by $C_{ij}$ and $T_{ij}$, respectively.
	The set of vehicles is represented by $V$. The load quantity change at location $i \in N$ is expressed as~$q_i$, with $q_0 = q_{2n+1} = 0$ and $q_i = -q_{n+i}$ ($q_i \in \mathbb{N}^+$) for $i \in P$. This quantity represents the customer size, such as the size of an order or the number of passengers. At pickup location $i \in P$, the vehicle adds load $q_i$, while at delivery location $i+n \in D$, the vehicle reduces load by $q_i$.
	
	The objective of DARP-LPT is to minimize travel costs while ensuring that at most $|V|$ vehicles are utilized to visit all pickup and delivery locations, starting from the origin depot location $0$ and ending at the destination depot location~$2n+1$. To achieve this, a constraint is imposed where each pickup location $i\in P$ must be visited by the same vehicle as its corresponding delivery location~$i+n\in D$ (pairing constraints), with the pickup location visited before the delivery location (precedence constraints). Moreover, each customer~$i\in P$ has a maximum ride time $R_i$, each location~$i\in N$ has its time window, and each vehicle is limited to a capacity of at most $Q$ with its maximum ride time $R_0$. 
	
	Up to this point, all the statements adhere to the standard DARP framework. However, an important and unique feature in DARP-LPT is that each vehicle is allowed to have at most $L$ number of pickups. Notably, $Q$ and $L$ are distinct and represent different constraints. $Q$ restricts the maximum size of concurrent customer loads within the vehicle, and the concurrent load can both increase and decrease throughout the trip. In contrast, $L$ limits the total number of pickups, which can only increase monotonically. 
	
	DARP-LPT is NP-hard. This can be proven by a reduction from the Capacitated Vehicle Routing Problem (CVRP). CVRP involves arranging a route plan using a fleet of vehicles to visit a set of customers at minimum cost. Each vehicle has a capacity denoted by $C$ \citep{Cornuejols1993,Achuthan2003}. To perform the reduction, in a DARP-LPT instance, we set the maximum number of pickups per trip $L$ equal to the vehicle capacity $C$ and merge each customer's pickup and delivery nodes into a single node. This creates a problem instance equivalent to CVRP, where each merged node represents a customer with unit size and each vehicle has the capacity $C$ equal to $L$. Since CVRP is known to be NP-hard when $C\ge 3$ \citep{Chen}, this reduction demonstrates that DARP-LPT is also NP-hard. MDARP-LPT can also be proven to be NP-hard using a reduction similar to DARP-LPT.

	\subsection{Arc-based formulation of DARP-LPT}\label{ABF}

	The arc-based formulation (ABF), as presented below, is the same as that of \citet{doi:10.1287/opre.1060.0283}, except for the constraints~(\ref{length13}).

	The decision variables are as follows: 
	\begin{table}[h]
		\footnotesize
		\centering
		\caption{Decision variables of arc-based formulation}
		\begin{tabular}{@{}ll@{}}
			\toprule
			Decision variables 						 &Definition                   \\\midrule
			$f_{ijv}$                                  & = 1, if the arc  $(i,j)\in A$ is traversed by vehicle $v\in V$; = 0, otherwise\\
			$t_{iv}$                                  & the departure time of vehicle $v$ at location $i\in N$\\
			$Q_{iv}$                                  & the load of vehicle $v$ after visiting $i\in N$\\
			\bottomrule
		\end{tabular}
		\label{table:notation3.1}
	\end{table}
	\allowdisplaybreaks
	\begin{align}
		&& \min\sum_{v\in V} \sum_{(i,j)\in A} C_{ij} f_{ijv}\label{DARPLobj1}\\
		s.t.
		&&\sum_{v\in V} \sum_{j\in N}f_{ijv}                  &= 1    &&i\in P\label{cover2}\\
		&&\sum_{j \in N}f_{ijv}  &=\sum_{j\in N}f_{\textcolor{black}{i+n,j,v}}      &&i\in P, v\in V  \label{pair3}\\
		&&\sum_{j\in N}f_{0jv}  &=1      &&v\in V  \label{flow4}\\
		&&\sum_{j\in N}f_{jiv}  &=\sum_{j\in N}f_{ijv}     &&i\in P\cup D, v\in V  \label{flow5}\\
		&&\sum_{i\in N}f_{\textcolor{black}{i,2n+1,v}}  &=1      &&v\in V  \label{flow6}\\
		&&t_{iv}+ T_{ij}  &\le t_{jv} +M_{ij}(1-f_{ijv})     &&(i,j)\in A, v\in V  \label{time7}\\
		&&e_i \le t_{iv} &\le l_i      &&i\in N,  v\in V  \label{time8}\\
		&&Q_{iv}+q_{j}  &\le Q_{jv} + W_{ij}(1-f_{ijv})     &&(i,j)\in A, v\in V  \label{capacity9}\\
		&&\max(0,q_i) \le Q_{iv} &\le \min(Q,Q+q_i)      &&i\in N,  v\in V  \label{capacity10}\\
		&&t_{\textcolor{black}{n+i,v}}- t_{iv} &\le R_i      &&i\in P, v\in V  \label{maxride11}\\
		&&t_{\textcolor{black}{2n+1,v}}- t_{0v} &\le R_0      && v\in V  \label{maxride12}\\
		&&\sum_{(i,j)\in A}  f_{ijv} 				&\le 2L+1		&&v\in V \label{length13}\\
		&&f_{ijv} 				&\in\{0, 1\}	&&(i,j)\in A, v\in V \label{domain}\\
		&&t_{iv}				&\ge 0	&&i\in N, v\in V \label{domain1}\\
		&&Q_{iv}				&\in\mathbb{N}	&&i\in N, v\in V \label{domain2}
	\end{align}	
	\allowdisplaybreaks
	
	The objective function (\ref{DARPLobj1}) aims to minimize the total cost of traversing arcs.
	Constraints~(\ref{cover2}) and (\ref{pair3}) ensure that each pickup location is visited exactly once and that the paired locations are traversed by the same vehicle.
	Constraints (\ref{flow4}) to (\ref{flow6}) specify the network flow requirements.
	Constraints (\ref{time7}) to (\ref{capacity10}) ensure time and load consistency. Here, $M_{ij}=\max(0, l_i+T_{ij} -e_j)$ and  $W_{ij}=\min(Q,Q+q_i)$.
	Constraints (\ref{maxride11}) and constraints (\ref{maxride12}) set the maximum ride time for customers and vehicles.
	Constraints~(\ref{length13}) restrict the maximum number of pickup locations per trip. If the number of pickups on a route is at most $L$, then the route has at most $2L$ pickup and delivery locations and $2L+1$ arcs.
	Constraints~(\ref{domain}) to (\ref{domain2}) specify the domain of the variables $f$, $t$, and $Q$.

	In order to reduce part of the symmetry that is inherent in the formulation, we define $V_f = \{1, 2, \ldots, \lceil n/L \rceil\}$ as the fixed set of vehicles that must be used. This definition arises from the observation that, if a vehicle can deliver at most $L$ customers, a minimum of $\lceil n/L \rceil$ vehicles are needed. Therefore, we add $f_{\textcolor{black}{0,2n+1,v}}=0$ for each fixed vehicle $v\in V_f$.

	\subsection{Fragment-based formulations of DARP-LPT}\label{fragmentformulate}

	The framework presented by \citet{Rist2021} defines a fragment-based formulation for DARP, which necessitates the inclusion of callbacks to handle rarely violated constraints related to subtour elimination, time windows, and maximum ride time. In the case of DARP, even when multiple vehicles are used, this fragment-based format remains feasible. 
	
	\textcolor{black}{This section builds upon the fragment-based formulations introduced by \citet{Rist2021}.   We propose two extended formulations to suitably manage the frequent violation of specific constraints in DARP-LPT, particularly the limited number of pickups per trip. Formulation FFF augments the decision variables by incorporating vehicle indices, while PSFF reformulates the network using a pickup-space structure and omits vehicle details. In the case of MDARP-LPT, FFF retains its structure, whereas PSFF must reintroduce vehicle indices to account for the unique origin and destination associated with each vehicle.}
	
	\subsubsection{Preliminaries and notation}\label{Preliminaries}
	
	In this subsection, we largely follow the definitions of \citet{Rist2021}.
	
	A \textit{DARP route} is a route $(0, i_1, i_2,\ldots, i_K, 2n+1)$ for which at least one feasible schedule exists, which respects the pairing, precedence, capacity, and time window constraints of DARP\@. The sequence of locations $(i_1, i_2, \ldots, i_K)$, excluding depot locations, is called a \textit{DARP route path}. Any segment of this sequence is referred to as a route path.

	\citet{Rist2021} introduce the concept of restricted fragments for DARP\@. These fragments can be combined to enumerate all DARP routes, and a network is constructed to include all these restricted fragments. Prior to introducing restricted fragments, we describe a full fragment in more detail.
	
	A \textit{full fragment} (FF) refers to a partial DARP route path where only the start and end nodes have an empty load. For instance, a valid fragment could have a sequence of locations $(p1,p2,d2,d1)$ or $(p1,p2,d1,p3,d2,d3)$, where $pj$ and $dj$ respectively denote the pickup and delivery locations for customer $j$. An invalid case is $(p1, d1, p2, p3, d2, d3)$ as this violates the empty load condition after location $d1$ and should be decomposed into $(p1, d1)$ and $(p2, p3, d2, d3)$. 
	
	A \textit{restricted fragment} (RF) is a part of a full fragment, starting at a pickup node whose predecessor is a delivery node or depot and ending at a delivery node whose successor is a pickup node or depot. 
	An RF contains \textcolor{black}{exactly} one pickup-to-delivery movement (not necessarily corresponding to a customer's pickup and delivery location pair), e.g., $(p1,p2,d1)$, $(p1,d2,d1)$, $(p1,p2,p3,d4,d5)$, $(p5,d8)$, $(p5,p7,d1,d2,d3)$. The following two examples are not RFs: $(p1, p2, p3, p4)$ and $(p2, d1, p3, d3, d2)$, where the former lacks a pickup-to-delivery movement and the latter can be split after $d1$.
	A special type of RF is a FF with the format ``$ppppdddd$'', where all pickups precede deliveries, and which cannot be divided into smaller RFs, for instance $(p1,p2,d1,d2)$ or $(p1,p3,p2,d2,d1,d3)$.
	Below are two further examples, where the left-hand side of the equality is the route path of an FF, and the right-hand side consists of RFs. This decomposition is entirely unique.
	
	$(p2,p1,d2,p4,d4,p3,d1,d3)= (p2,p1,d2)+(p4,d4)+(p3,d1,d3)$; 
	
	$(p2,p5,d2,p4,d4,p3,d5,d3)= (p2,p5,d2)+(p4,d4)+(p3,d5,d3)$.
	
	When combining RFs into routes, it will be crucial to track loads on board. In the first example, the load after visiting locations $d2$ and $d4$ is ${l_1}$, while in the second example, the load after visiting these locations is ${l_5}$, where $l_j$ denotes the load for customer $j$. This difference highlights that both the route path and the load information are essential for accurately representing an RF\@. A more precise definition for fragments (RF and FF) will be provided after the introduction of the following concepts.
	
	A \textit{node} is a pair consisting of a location and a set of loads (with customer identification and not just the number of customers) on board, denoted as $(loc(i), loadset(i))$ (or just $loc(i)$ if $loadset(i)=\emptyset$). The $loadset(i)$ includes the unserved customers along a path after departing from location $loc(i)$ but excludes those from $loc(i)$. For instance, the nodes from the route path $(p1, p2, d1, p3, d2, p4, d3, d4)$ are expressed as $p1$, $(p2,\{l_1\})$, $(d1,\{l_2\})$, $(p3,\{l_2\})$, $(d2,\{l_3\})$, $(p4,\{l_3\})$, $(d3,\{l_4\})$, and $d4$. Notably, nodes with the same location can be distinct, \textcolor{black}{as illustrated by examples such as $(p2,\{l_1\})$, $(p2,\{l_3\})$, $(p2,\{l_1, l_3\})$, and $p2$.} We define a node with a pickup location as a ``pickup node'' and a node with a delivery location as a ``delivery node''. Specifically, we denote the sets of pickup and delivery nodes as $P_N$~and~$D_N$, respectively, distinguishing them from the sets of pickup and delivery locations denoted as $P$~and~$D$. 
	
	A \textit{fragment} (FF and RF) is defined as a partial DARP route path with distinct start and end nodes. The route path records the sequence of locations traversed by a fragment, while the start and end nodes include information regarding the loads onboard. For a FF, merely the route path suffices for reference, while for an RF, alongside the route path, inclusion of both the start and end nodes is necessary for identification. For illustrative purposes, we present several example fragments: $p4-(p4,d4)-d4$,  $(p4,\{l_1\})-(p4,d4)-(d4,\{l_1\})$, $(p4,\{l_2\})-(p4,d4)-(d4,\{l_2\})$, $(p4,\{l_1, l_2\})-(p4,d4)-(d4,\{l_1, l_2\})$. In this context and below, for such a format, the first and last element represent the start and end nodes of the fragment respectively, while the part between two hyphens denotes the route path. Despite following the same route path, these fragments are not the same. 
	
	\subsubsection{Fragment-based network}

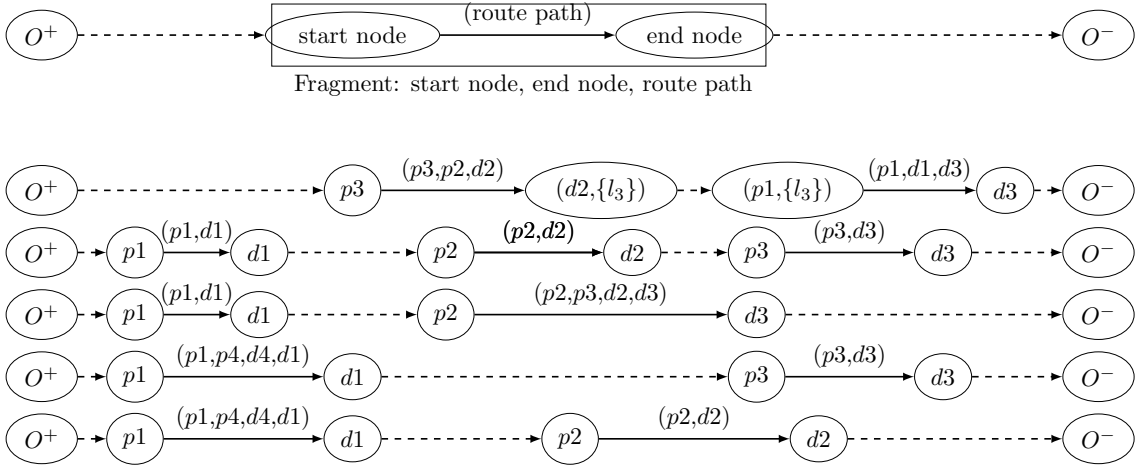
\begin{figure}[h]
		\centering
		\resizebox{0.99\textwidth}{!}{
			\begin{tikzpicture}
				\node[draw,ellipse] (0+) at (-5.5,1) {$O^+$};
				\node[draw, ellipse] (p3) at (-0.5,1) {$p3$};
				\node[draw, ellipse] (d2p3) at (3.5,1) {($d2$,\{$l_3$\})};
				\node[draw, ellipse] (p1p3) at (6.5,1) {($p1$,\{$l_3$\})};
				\node[draw, ellipse] (d3) at (10,1) {$d3$};
				\node[draw, ellipse] (0-) at (11.5,1) {$O^-$};
				
				\draw[line width=0.75pt, ->,dashed,>=latex] (0+) -> (p3);
				\draw[line width=0.75pt, ->,dashed,>=latex] (d3) ->  (0-);
				\draw[line width=0.75pt, ->,dashed,>=latex] (d2p3) -> (p1p3);
				
				\draw[line width=0.75pt, ->,>=latex] (p3) -- node[above] {($p3$,$p2$,$d2$)} (d2p3);
				\draw[line width=0.75pt, ->,>=latex] (p1p3) -- node[above] {($p1$,$d1$,$d3$)} (d3);

				\node[draw,ellipse] (10+) at (-5.5,0) {$O^+$};
				\node[draw, ellipse] (1p1) at (-4,0) {$p1$};
				\node[draw, ellipse] (1d1) at (-2,0) {$d1$};
				\node[draw, ellipse] (1p2) at (1,0) {$p2$};
				\node[draw, ellipse] (1d2) at (4,0) {$d2$};
				\node[draw, ellipse] (1p3) at (6,0) {$p3$};
				\node[draw, ellipse] (1d3) at (9,0) {$d3$};
				\node[draw, ellipse] (10-) at (11.5,0) {$O^-$};
				
				\draw[line width=0.75pt, ->,dashed,>=latex] (10+) -> (1p1);
				\draw[line width=0.75pt, ->,dashed,>=latex] (1d3) ->  (10-);
				\draw[line width=0.75pt, ->,dashed,>=latex] (1d1) -> (1p2);
				\draw[line width=0.75pt, ->,dashed,>=latex] (1d2) -> (1p3);
				
				\draw[line width=0.75pt, ->,>=latex] (1p1) -- node[above] {($p1$,$d1$)} (1d1);
				\draw[line width=0.75pt, ->,>=latex] (1p2) -- node[above] {($p2$,$d2$)} (1d2);
				\draw[line width=0.75pt, ->,>=latex] (1p3) -- node[above] {($p3$,$d3$)} (1d3);

				\node[draw,ellipse] (10+) at (-5.5,-1) {$O^+$};
				\node[draw, ellipse] (1p1) at (-4,-1) {$p1$};
				\node[draw, ellipse] (1d1) at (-2,-1) {$d1$};
				\node[draw, ellipse] (11p2) at (1,-1) {$p2$};
				\node[draw, ellipse] (11d3) at (6,-1) {$d3$};
				\node[draw, ellipse] (10-) at (11.5,-1) {$O^-$};
				
				\draw[line width=0.75pt, ->,dashed,>=latex] (10+) -> (1p1);
				\draw[line width=0.75pt, ->,dashed,>=latex] (11d3) ->  (10-);
				\draw[line width=0.75pt, ->,dashed,>=latex] (1d1) -> (11p2);
				
				\draw[line width=0.75pt, ->,>=latex] (1p1) -- node[above] {($p1$,$d1$)} (1d1);
				\draw[line width=0.75pt, ->,>=latex] (11p2) -- node[above] {($p2$,$p3$,$d2$,$d3$)} (11d3);
				
				\node[draw,ellipse] (10+) at (-5.5,-2) {$O^+$};
				\node[draw, ellipse] (1p1) at (-4,-2) {$p1$};
				\node[draw, ellipse] (1d1) at (-0.5,-2) {$d1$};
				\node[draw, ellipse] (1p3) at (6,-2) {$p3$};
				\node[draw, ellipse] (1d3) at (9,-2) {$d3$};
				\node[draw, ellipse] (10-) at (11.5,-2) {$O^-$};
				
				\draw[line width=0.75pt, ->,dashed,>=latex] (10+) -> (1p1);
				\draw[line width=0.75pt, ->,dashed,>=latex] (1d3) ->  (10-);
				\draw[line width=0.75pt, ->,dashed,>=latex] (1d1) -> (1p3);
				
				\draw[line width=0.75pt, ->,>=latex] (1p1) -- node[above] {($p1$,$p4$,$d4$,$d1$)} (1d1);
				\draw[line width=0.75pt, ->,>=latex] (1p2) -- node[above] {($p2$,$d2$)} (1d2);
				\draw[line width=0.75pt, ->,>=latex] (1p3) -- node[above] {($p3$,$d3$)} (1d3);

				\node[draw,ellipse] (10+) at (-5.5,-3) {$O^+$};
				\node[draw, ellipse] (1p1) at (-4,-3) {$p1$};
				\node[draw, ellipse] (1d1) at (-0.5,-3) {$d1$};
				\node[draw, ellipse] (1p2) at (3,-3) {$p2$};
				\node[draw, ellipse] (1d2) at (7,-3) {$d2$};
				\node[draw, ellipse] (40-) at (11.5,-3) {$O^-$};
				
				\draw[line width=0.75pt, ->,dashed,>=latex] (10+) -> (1p1);
				\draw[line width=0.75pt, ->,dashed,>=latex] (1d2) ->  (40-);
				\draw[line width=0.75pt, ->,dashed,>=latex] (1d1) -> (1p2);
				
				\draw[line width=0.75pt, ->,>=latex] (1p1) -- node[above] {($p1$,$p4$,$d4$,$d1$)} (1d1);
				\draw[line width=0.75pt, ->,>=latex] (1p2) -- node[above] {($p2$,$d2$)}(1d2);
				
				\node[draw,ellipse] (0+) at (-5.5,3.0) {$O^+$};
				\node[draw, ellipse] (6p3) at (-0.5,3.0) {start node};
				\node[draw, ellipse] (6d3) at (5,3.0) {end node};
				\node[draw, ellipse] (0-) at (11.5,3.0) {$O^-$};
				
				\draw[line width=0.75pt, ->,dashed,>=latex] (0+) -> (6p3);
				\draw[line width=0.75pt, ->,dashed,>=latex] (6d3) ->  (0-);
				
				\draw[line width=0.75pt, ->,>=latex] (6p3) -- node[above] {(route path)} (6d3);

				\draw (-1.9,2.5) rectangle (6.26,3.6);
				\node[below] at (2.25,2.5) {Fragment: start node, end node, route path};
		\end{tikzpicture}}
		\caption{Several feasible routes for DARP and their decomposition into fragments (RF and FF)}\label{fig1}
	\end{figure}

	Building upon the introductory concepts, we now introduce the fragment-based network. \textcolor{black}{To facilitate comprehension, we provide Fig.~\ref{fig1} and Fig.~\ref{fig3}. Fig.~\ref{fig1} illustrates several DARP routes alongside their corresponding fragments, whereas Fig.~\ref{fig3} depicts the corresponding fragment-based network. This network can be derived from Fig.~\ref{fig1} by first merging all identical fragments (with an intermediate transformation shown in Fig.~\ref{fig2} in \ref{AEFBN}), and then merging identical nodes while preserving the original connections.} In these figures, ellipses represent nodes, solid lines (with ellipses on both sides) indicate fragments (with their route paths written above or to the right), and dashed lines with arrows are node arcs (further explained below). $O^+$ and $O^-$ denote the origin depot and destination depot, respectively.

	\begin{figure}[h]
		\centering
		\resizebox{0.75\textwidth}{!}{
			\begin{tikzpicture}
				\node[draw,ellipse] (0+) at (-8,1) {$O^+$};
				\node[draw, ellipse] (0-) at (5,1) {$O^-$};
				\node[draw, ellipse] (p2) at (-0.5,1) {$p2$};
				\node[draw, ellipse] (p3) at (-3,3) {$p3$};
				\node[draw, ellipse] (p1) at (-6,1) {$p1$};
				\node[draw, ellipse] (d2) at (2.5,1) {$d2$};
				\node[draw, ellipse] (d3) at (3,3) {$d3$};
				\node[draw, ellipse] (d1) at (-2,1) {$d1$};
				\node[draw, ellipse] (d2p3) at (-1.5,5) {($d2$,\{$l_3$\})};
				\node[draw, ellipse] (p1p3) at (2,5) {($p1$,\{$l_3$\})};
				
				\draw[line width=0.75pt, ->,dashed,>=latex] (0+) to[out=40, in=-180]  (p3);
				\draw[line width=0.75pt, ->,dashed,>=latex] (0+) ->  (p1);
				\draw[line width=0.75pt, ->,dashed,>=latex] (d2) ->  (0-);
				\draw[line width=0.75pt, ->,dashed,>=latex] (d3) ->  (0-);
				\draw[line width=0.75pt, ->,dashed,>=latex] (d2)  to[out=140, in=-10]  (p3);
				\draw[line width=0.75pt, ->,dashed,>=latex] (d2p3) ->  (p1p3);
				\draw[line width=0.75pt, ->,dashed,>=latex] (d1) -> (p3);
				\draw[line width=0.75pt, ->,dashed,>=latex] (d1) -> (p2);

				\draw[line width=0.75pt, ->,>=latex] (p2) -- node[above] {($p2$,$d2$)} (d2);
				\draw[line width=0.75pt, ->,>=latex] (p3) -- node[above] {($p3$,$d3$)} (d3);
				\draw[line width=0.75pt, ->,>=latex] (p1) -- node[above] {($p1$,$d1$)} (d1);
				\draw[line width=0.75pt, ->,>=latex] (p1) to[out=40, in=140] node[above] {($p1$,$p4$,$d4$,$d1$)} (d1);
				\draw[line width=0.75pt, ->,>=latex] (p3)-- node[right] {($p3$,$p2$,$d2$)} (d2p3);
				\draw[line width=0.75pt, ->,>=latex] (p1p3)-- node[right] {($p1$,$d1$,$d3$)} (d3);
				\draw[line width=0.75pt, ->,>=latex] (p2)  to[out=40, in=-165] node[right]{($p2$,$p3$,$d2$,$d3$)} (d3);

		\end{tikzpicture}}
		\caption{Fragment-based network for DARP by merging all identical \textcolor{black}{fragments and} nodes from \textcolor{black}{Fig.~\ref{fig1}}}\label{fig3}
	\end{figure}
	
	Before explaining the fragment-based network, we will introduce the \textit{connections between fragments} and the \textit{connections between nodes} (referred to later as node arcs), both of which can be seen as the dashed lines in Fig.~\ref{fig1}. In the former category, the connections between FFs are straightforward. For example:
	$p1-(p1, d1)-d1$ can connect to $p2-(p2, p3, d2, d3)-d3$ to form a route path $(p1, d1, p2, p3, d2, d3)$. 
	However, the connections between RFs require ensuring consistency between the last node's load of the preceding fragment and the first node's load of the succeeding one. Take the following three RFs as an example.\\
	$p2-(p2,p1,d2)-(d2,\{l_1\})$,\\
	$(p4,\{l_1\})-(p4, d1)-(d1,\{l_4\})$,\\
	$(p3,\{l_4\})-(p3, d4, d3)-d3$.\\
	These can be combined to a route path $(p2,p1,d2,p4,d1,p3,d4,d3)$. However, the first RF cannot connect to the third one since the load of the end node of the first RF is not the same as that of the start node of the third one. 
	
	The connection between two fragments can be indicated by the connection between the preceding fragment's end node and the succeeding fragment's start node. For instance, in the third row of Fig.~\ref{fig1}, the connection between $p1-(p1, d1)-d1$ and $p2-(p2, d2)-d2$ can be regarded as the connection between $d1$ and $p2$. Furthermore, the connections between distinct fragments sharing identical end nodes~$h^-$ and different fragments with identical start nodes~$h^+$, can be consistently depicted by the connection $(h^-,h^+)$. For instance, in the third and fourth row of Fig.~\ref{fig1}, the connections between $p1-(p1, d1)-d1$ and $p2-(p2, d2)-d2$, as well as between $p1-(p1, d1)-d1$ and $p2-(p2, p3, d2, d3)-d3$, can both be represented by the connection $(d1,p2)$.
	
	Therefore, we introduce the concept of \textit{node arc} to describe such a connection between nodes. In the node arc $(h^-, h^+)$, $h^-$ and $h^+$ then represent the start and end nodes of the node arc, respectively. This node arc concept differs from the arc concept defined in Section~\ref{ABF}, as node arcs connect one node to another, with each node encompassing both location and load information (unless the node has empty loads). An additional note is that a node arc can only represent a connection from a delivery node or the origin depot to a pickup node or from a delivery node to the destination depot. This is because all fragments' end nodes must be delivery nodes, while their start nodes must be pickup nodes. Crucially, the connected nodes must share the same load.

	The practical process of network generation proceeds as follows.
	The fragment-based network comprises nodes, fragments, and node arcs. The initial step involves generating fragments (outlined in Section~\ref{fragmethod}). Following this, the start and end nodes of all fragments are identified and added to the network. Subsequently, all fragments are integrated into the network, connecting from their start nodes to end nodes. Different fragments can share the same start and/or end nodes. In the final phase, for the node arcs, all delivery nodes and the origin depot connect with all pickup nodes, and all delivery nodes connect with the destination depot. If none of the fragments ending at $h^-$ can connect to any fragment starting at $h^+$ due to time window or any other constraints, the node arc $(h^-,h^+)$ can be eliminated. When the network exclusively includes FFs, the number of node arcs (excluding those involving depots) is at most $|n| \times |n|$. For the network with RFs, the potential number of node arcs could be higher due to the possibility of different nodes with the same location having different loads.
	Moreover, a feasible route typically starts with a node arc from the depot, followed by a fragment leading to a delivery node. Subsequently, there must be a node arc from the current delivery node to another pickup node or the destination depot. After that, if the current node is not the depot yet, it must be succeeded by another fragment and then a node arc again, continuing in this pattern. Therefore, any feasible route in Fig.~\ref{fig3} must adhere to the alternating pattern of dashed lines and solid lines. \textcolor{black}{For details on why Fig.~\ref{fig3} can be used to enumerate all paths, please refer to \ref{AEFBN}.}

	We now formalize the foregoing concepts. The set of \textit{nodes} is expressed as: $N_N=P_N\cup D_N\cup\{ O^+, O^-\}$, where the origin depot $O^+$ and destination depot $O^-$ correspond to location $0$ and location $2n+1$, respectively. The set of fragments, denoted as $F$, is derived from customer locations, as detailed in Section~\ref{fragmethod}. $F$ may include both FFs and RFs or exclusively one type. We use the notation $F_h^+$ for the sets of fragments starting at pickup node $h\in P_N$, $F_h^-$ for those ending at delivery node $h\in D_N$, and $F_i$ for the set of fragments traversing location $i\in P\cup D$. Additionally, $A_N$, $A^+_n$, and $A^-_n$ represent the set of node arcs, the set of node arcs starting from delivery node (or the origin depot) $h\in D_N\cup \{O^+\}$, and the set of node arcs ending at pickup nodes (or the destination depot) $h\in P_N\cup \{O^-\}$, respectively. Notably, $A_N$ exclusively includes node arcs originating from delivery nodes (or the origin depot) and terminating at pickup nodes (or the destination depot).
	
	For each fragment $f\in F$, we designate the number of included pickup nodes as $L_f$, and its cost $c_f$ is calculated as the sum of travel costs over all locations visited by the fragment, i.e., $c_f=\sum_{(i,j)\in f} C_{loc(i)loc(j)}$, where $loc(i)$ represents the location of node $i\in N_N$. Furthermore, the cost of node arc $a$, $c_a$, is determined as follows: $c_a=C_{loc(i)loc(j)}$, $a=(i,j)$, $a\in A_N$.
	
	\subsubsection{Fragment flow formulation}\label{FFFormulation}
	
	The first fragment-based formulation is the \textit{fragment flow formulation} (FFF), which expands on the framework of \citet{Rist2021}. It transitions from a formulation using fragment-based and node arc-based decision variables to one employing fragment-vehicle-based and node arc-vehicle-based variables.
	Please note that the current formulation below is not complete. Lazy constraints, specifically designed for the elimination of infeasible paths, are applied to address subtour elimination, time windows, and maximum ride time. The details are described in Section~\ref{inequalities}. 
	
	\begin{table}[H]
		\footnotesize
		\centering
		\caption{Decision variables of FFF}
		\begin{tabular}{@{}ll@{}}
			\toprule
			Decision variables 						 &Definition                   \\\midrule
			$x_{fv}$& = 1, if fragment  $f\in F$ is traversed by vehicle $v\in V$; = 0, otherwise\\
			$y_{av}$& = 1, if node arc $a\in A_N$ is traversed by vehicle $v\in V$; = 0, otherwise  \\\bottomrule
		\end{tabular}
		\label{table:notation3.2}
	\end{table}
	
	\begin{align}
		\min \sum_{v\in V} \sum_{f\in F} c_{f} x_{fv}+ \sum_{v\in V} \sum_{a\in A_N} c_{a} y_{av}\label{MDARPobj1}
	\end{align}	
	\begin{align}
		s.t.
		&&\sum_{f\in F_h^-}x_{fv}  &=\sum_{a\in A_h^+}y_{av}      &&h\in P_N, v\in V  \label{FragmentArc1}\\
		&&\sum_{f\in F_h^+}x_{fv}  &=\sum_{a\in A_h^-}y_{av}      &&h\in D_N, v\in V  \label{ArcFragment1}\\
		&&\sum_{v\in V} \sum_{f\in F_i} x_{fv}                  &= 1    &&i\in P\label{cover1}\\
		&&\sum_{f\in F}L_f x_{fv}                  &\le L    &&v\in V\label{length}\\
		&&x_{fv}, y_{av}					&\in\{0, 1\}		&&f{\in} F, a{\in}A_N, v{\in}V\label{DARdomain1}
	\end{align}	
	
	The objective function~(\ref{MDARPobj1}) minimizes the weighted cost of used fragments and node arcs.
	Constraints~(\ref{FragmentArc1}) and (\ref{ArcFragment1}) describe the network flow for fragments and node arcs. For each pickup node at which a fragment ends, there must be a node arc starting from it. Similarly, for each delivery node at which a fragment starts, a node arc must end at that node. These relationships are applicable to each vehicle.
	Constraints~(\ref{cover1}) guarantee that each pickup location is visited exactly once.
	Constraints~(\ref{length}) set the maximal number of pickups.
	Constraints~(\ref{DARdomain1}) specify the domains of $x$ and $y$.
	
	The additional constraints~(\ref{vehicle}) below specify whether a vehicle is used or unused. They are not necessary but can assist the solver in branching, and thus, improve computational performance. Here, we define $w_{v}$ as 1 if vehicle $v \in V$ is utilized, and 0 otherwise.
	
	\begin{align}
		&&	\sum_{a\in A_{O^+}^+} y_{av}                  &= w_v   &&v\in V\label{vehicle}
	\end{align}

	\textcolor{black}{Furthermore, to enhance the formulation FFF, we introduce additional constraints based on the minimum number of used vehicles, as shown in constraints~(\ref{fixvehilceused1}). While these constraints may be redundant, they significantly speed up the computation process. Additionally, for constraints~(\ref{vehicle}), we set $w_v$ equal to 1 for $v \in V_f$.}
	
	\begin{align}
		&& \sum_{f\in F}x_{vf}                &\ge 1    &&v\in V_f\label{fixvehilceused1}
	\end{align}
	
	\subsubsection{\textcolor{black}{Pickup-space fragment formulation}}\label{PSFFormulation}
	
	\textcolor{black}{We now describe the Pickup-Space Fragment Formulation (PSFF), which extends the fragment-based formulation of \citet{Rist2021} by introducing the number of pickups per trip as a new dimension, analogous to a time-space representation. Unlike FFF, which uses a separate constraint for pickup limits, PSFF integrates this constraint directly into the pickup-space network.}
	
	\textcolor{black}{The pickup-space network is formed by replicating the fragment-based network across $L+1$ levels, where each level represents the remaining number of allowable pickups throughout a trip. The index sets $\mathcal{L_P} = \{1, \dots, L\}$ and $\mathcal{L_D} = \{0, \dots, L-1\}$ define the remaining pickup levels for pickup and delivery nodes, respectively. When a vehicle traverses a fragment with $l \in \mathcal{L_P}$ remaining allowable pickups before entry, it concludes at level $l - L_f \in \mathcal{L_D}$. The transitions for node arcs adhere to the same level structure, ensuring consistency in tracking the remaining allowable number of pickups.}
	
	\textcolor{black}{The following formulation does not include all constraints. Lazy constraints are incorporated to address subtour elimination, time windows, and ride-time constraints, as detailed in Section~\ref{inequalities}.}
	
	\begin{table}[H]\textcolor{black}{
			\footnotesize
			\centering
			\caption{Decision variables of LSF}
			\begin{tabular}{@{}ll@{}}
				\toprule
				Decision Variables & Definition \\ \midrule
				$X_{fl}$ & = 1, if a vehicle traverses fragment $f \in F$ with $l$ remaining allowable pickups \\ & before entry; = 0, otherwise \\
				$Y_{al}$ & = 1, if a vehicle traverses node arc $a \in A_N$ while having $l$ remaining allowable \\ & pickups available; = 0, otherwise \\
				\bottomrule
			\end{tabular}
			\label{table:decisionvariablesLSF}
	}\end{table}

	\textcolor{black}{	\begin{align}
			\min \sum_{l \in \mathcal{L_P}}\sum_{f \in F} c_f X_{fl} + \sum_{l \in \mathcal{L_D}}\sum_{a \in A_N} c_a Y_{al} \label{PFF_obj}
		\end{align}
		\begin{align}
			s.t. \quad 
			&& \sum_{f \in F_h^-} X_{f, l + L_f} &= \sum_{a \in A_h^+} Y_{a l} &&  h \in P_N, \; l \in \mathcal{L_P} \label{flow_pickup} \\
			&& \sum_{f \in F_h^+} X_{f l} &= \sum_{a \in A_h^-} Y_{al} && h \in D_N, \; l \in \mathcal{L_D} \label{flow_delivery} \\
			&& \sum_{l \in \mathcal{L_P}} \sum_{f \in F_i} X_{f l} &= 1 &&  i \in P_N \label{covering} \\
			&& \sum_{a \in A_o^{+}} Y_{a L} &\leq |V| && \label{vehicle_limit} \\
			&& X_{f l} &\in \{0,1\} &&  f \in F \; l \in \mathcal{L_P} \label{domain_PFFX} \\
			&& Y_{a l} &\in \{0,1\} &&  a \in A_N \; l \in \mathcal{L_D} \label{domain_PFFY}
	\end{align}}
	\textcolor{black}{The objective~\ref{PFF_obj} is to minimize the total weighted cost of fragments and arcs. Constraints~(\ref{flow_pickup}) and (\ref{flow_delivery}) establish the network flow constraints on each pickup and delivery node at every remaining pickup level. These constraints help track the remaining allowable pickup number throughout the trip. Specifically, constraints~(\ref{flow_pickup}) ensure that if a fragment~$f$ initially has a remaining allowable pickup number of $l + L_f$ then it will end at a delivery node with level~$l$, and then a node arc must depart from this delivery node at the same level. Constraints~(\ref{flow_delivery}) specify that if a fragment departs from a pickup node at level~$l$, then a node arc must end at this pickup node at level~$l$ (and this node arc also starts at level~$l$). Constraint~(\ref{covering}) ensures that each pickup location is visited exactly once, thereby guaranteeing that all customer requests are fulfilled. Constraint~(\ref{vehicle_limit}) imposes an upper bound on the total number of vehicles used in the network. In particular, it restricts the number of pickup-space node arcs that depart from the origin depot with a remaining number of pickups equal to $L$. Constraints~(\ref{domain_PFFX}) and (\ref{domain_PFFY}) define the domains of $X$ and $Y$.}

	\subsubsection{Inequalities for fragment formulations}\label{inequalities}
	
	The above fragment-based formulations FFF and \textcolor{black}{PSFF} may not be sufficient to lead to feasible solutions. Subtour elimination, time windows, and maximum ride time constraints will be implemented by the callbacks of infeasible path constraints, in line with \citet{Alyasiry2019, Rist2021}. The details are as follows. As long as we encounter an incumbent solution in the branch-and-bound tree, we create chains or cycles. A \textit{chain} is a sequence of fragments $(f_1, \dots,f_c)$ with a fragment count of $c>1$, along with $c-1$ node arcs $(a_1, \dots,a_{c-1})$ connecting them. The load of the end node of fragment $f_i$ must be the same as that of the start node of $f_{i+1}$ for $1\le i \le c-1$; the start node and the end node of $a_i$ correspond to the end node of fragment $f_i$ and the start node of $f_{i+1}$, respectively, for $1\le i \le c-1$. Similarly, a \textit{cycle} is a sequence of fragments $(f_1, \dots,f_c)$ with a fragment count of $c>1$, along with $c$ node arcs $(a_1, \dots,a_{c})$ connecting them. A cycle has the same connection requirement as a chain but includes an additional node arc $a_c$ that connects $f_c$ back to $f_1$.
	
	\textcolor{black}{For FFF,} if we obtain a chain, we connect its initial and final fragments to the origin and destination depots, respectively, to construct a path. After that, we establish a valid time schedule for the path, following the constraints of time windows and maximum ride time. If no valid schedule can be found, the chain is considered infeasible and then needs to be eliminated by constraints~(\ref{infeasiblePath}), which ensures that the total count of fragments and node arcs in such a chain does not exceed $2c-2$. If a cycle is detected, we then remove it using constraints~(\ref{infeasiblecycle}) to ensure that the total count of its elements does not exceed $2c-1$, thereby achieving subtour elimination. While callbacks are a useful approach for rarely violated constraints, like the constraints mentioned above, they are not ideal for constraints that are easily violated, such as the limit on the number of pickups. 
	\begin{align}
		&&\sum_{k=1}^c x_{f_kv} + \sum_{k=1}^{c-1} y_{a_kv}&\le 2c-2&&v\in V\label{infeasiblePath} \\
		&&\sum_{k=1}^c x_{f_kv} + \sum_{k=1}^{c} y_{a_kv}&\le 2c-1&&v\in V\label{infeasiblecycle}
	\end{align}
	
	\textcolor{black}{For PSFF the procedure remains the same, but for the callback constraints the variables $x$ and $y$ in constraints~(\ref{infeasiblePath}) and~(\ref{infeasiblecycle}) are replaced by $\sum_{l \in \mathcal{L_P}} X_{f_k l}$ and $\sum_{l \in \mathcal{L_D}} Y_{a_k l}$, respectively, and the vehicle index and vehicle set are no longer required.}
	

	\subsection{\textcolor{black}{Path-based formulation}}\label{PBF}
	
	\textcolor{black}{Since a fragment generally spans a longer sequence than a location arc, its enumeration requires more computational effort. To further understand the trade-off, we enumerate all feasible paths and solve the problem using a path-based formulation (PBF). Given the limited number of pickups per trip, such enumeration may become practically viable, especially when the maximum number of pickups $L$ is small or when the instance size is relatively small.}
	
	\textcolor{black}{The path-based formulation follows a set partitioning structure, as in \citet{Ropke2007}. Let $\omega$ denote the set of all enumerated feasible paths. For each path $p \in \omega$, we define $a_p^i$ as a binary parameter that takes value 1 if location~$i \in N$ is visited in path $p$, and 0 otherwise. Let $c_p$ represent the cost associated with path $p$. The decision variable~$z_p$ equals 1 if path $p$ is selected, and 0 otherwise. The formulation is given below.}
	\textcolor{black}{	\begin{align}
			\min \quad & \sum_{p \in \omega} c_p z_p \label{path_obj} \\
			s.t.\quad & \sum_{p \in \omega} a_p^i z_p = 1 &&  i \in N \label{path_covering_constraint} \\
			&\sum_{p \in \omega} z_p \le |V| && \label{path_maxvehicles}\\
			& z_p \in \{0,1\} &&  p \in \omega \label{path_binary}
	\end{align}}
	\textcolor{black}{The objective~\eqref{path_obj} minimizes the total cost of the selected paths. Constraints~\eqref{path_covering_constraint} ensure that each location is visited exactly once, while constraint~\eqref{path_maxvehicles} specifies the maximum number of vehicles. Finally, constraints~\eqref{path_binary} define the binary domain of the decision variables $z$.}

	\subsection{MDARP-LPT: problem description and formulations}\label{MDARPProbForm}
	
	The primary distinction between MDARP-LPT and DARP-LPT lies in the fact that in DARP-LPT all drivers must initiate their journeys from the origin depot $O^+$ and return to the destination depot $O^-$. Conversely, in MDARP-LPT, each vehicle $v\in V$ possesses its own unique origin location $o_v$ and destination location $d_v$ (denoted as $O_v$ and $D_v$ for origin and destination nodes, respectively). In all other aspects, MDARP-LPT adheres to the same principles as DARP-LPT. 
	
	The formulations developed supra (ABF, FFF, \textcolor{black}{PSFF, and PBF}) for DARP-LPT can also be used for MDARP-LPT, with some modifications as explained below.
	
	For ABF, only the network flow constraints~(\ref{flow4})-(\ref{flow6}) need to be updated to reflect the unique origin and destination for each vehicle. \textcolor{black}{Specifically, the generic locations $0$ and $2n+1$ are updated to $o_v$ and $d_v$, respectively.
		For the case of FFF and PSFF, all constraints remain applicable, but the network configuration assigns unique origin and destination nodes, fragment sets, node arc sets, and node sets to each vehicle. Finally, for PBF the only modification is to replace the constraint~\eqref{path_maxvehicles} by a separate inequality $\sum_{p \in \omega_{v}} z_p \le 1$ for each vehicle~$v$, where $\omega_{v}$ denotes the set of paths associated with vehicle~$v$.}

	\section{Fragment generation}\label{fragmethod}
	
	\textcolor{black}{This section describes how the fragments that constitute the basis for the formulations FFF and PSFF can be acquired. Section~\ref{tradFrag} outlines the enumeration-based fragment generation procedures of FFs \citep{Alyasiry2019} and RFs \citep{Rist2021}, resulting in fragment sets $S_{FF}$ and $S_{RF}$, respectively. Section~\ref{exFrag} introduces extended fragments (EFF) and extended restricted fragments (ERF), along with their generation procedures for the corresponding sets $S_{EFF}$ and $S_{ERF}$. Finally, Section~\ref{fragmentselect} describes a novel mixed fragment set $S_{MF}$, derived by converting FFs from $S_{FF}$ into RFs and ERFs. This approach aims to reduce the fragment count while maintaining more useful fragments.}
	
	From these sets, all fragments whose length exceeds $L$ are eliminated. The set $S_{FF}$, containing longer fragments, tends to be larger than $S_{RF}$, which contains shorter fragments. However, all combinations of RFs from $S_{RF}$ can generate more (redundant) FFs than those present in $S_{FF}$.  \textcolor{black}{This observation also applies to the comparison between $S_{ERF}$ and $S_{EFF}$.} 
	

	All these fragment sets \textcolor{black}{($S_{FF}$, $S_{RF}$, $S_{EFF}$, $S_{ERF}$, and $S_{MF}$)} are valid for both MDARP-LPT and DARP-LPT\@. However, in MDARP-LPT, where vehicles have different origins and destinations, each vehicle requires a unique fragment set (though many of the fragments will still be the same for each vehicle). Therefore, the generation procedure for MDARP-LPT must be executed $|V|$ times, where $|V|$ denotes the number of vehicles. In contrast, DARP-LPT, with a fixed depot for all vehicles, only necessitates a single execution of the generation procedure.
	
	\subsection{\textcolor{black}{RF and FF generation}}\label{tradFrag}
	
	\citet{Alyasiry2019} introduce FFs and $S_{FF}$, while \citet{Rist2021} propose RFs and $S_{RF}$. $S_{FF}$ exclusively contains FFs, while $S_{RF}$ exclusively contains RFs. Their intersection \textcolor{black}{is} all FFs with the format ``$ppppdddd$'', which can also be regarded as RFs. As previously explained, ``$ppppdddd$'' indicates a sequence where all pickups precede all deliveries.
	
	The fragment generation procedure of \citet{Alyasiry2019} involves enumerating all FFs that satisfy constraints such as time windows, maximum ride time, pairing, precedence, and capacity limits. In \citet{Rist2021}, fragment generation is relatively intricate. Although an RF is always a part of at least one FF (see Section~\ref{Preliminaries}), \citeauthor{Rist2021} do not directly derive RFs from FFs. Instead, they only focus on FFs with the specific format ``$ppppdddd$''. These specific FFs are initially enumerated and \textcolor{black}{all possible substrings are then generated}. For example, ($p1,p2,p3,d1,d2,d3$) \textcolor{black}{leads to the following list of sequences}:
	($p1,p2,p3,d1,d2,d3$),
	($p1,p2,p3,d1,d2$),
	($p1,p2,p3,d1$),
	($p2,p3,d1,d2,d3$),
	($p2,p3,d1,d2$),
	($p2,p3,d1$),
	($p3,d1,d2,d3$),
	($p3,d1,d2$),
	($p3,d1$).
	Building upon \textcolor{black}{these} results, \textcolor{black}{\citet{Rist2021}} then systematically enumerate all feasible RFs. An additional step involves transforming a \textcolor{black}{generated sequence} above into an RF\@. For each segment, the left and right parts (customers) cut from the original FF would contribute to the load set of the start and end nodes of the newly constructed RF, respectively. For example, for the segment ($p3, d1$), its left side ($\textbf{p1, p2}$) and right side ($\textbf{d2, d3}$) were cut from the original FF ($\textbf{p1,p2},p3,d1,\textbf{d2,d3}$); then for the newly formed RF, the start node becomes ($p3, \{l_1, l_2\}$) and the end node becomes ($d1, \{l_2, l_3\}$). Hence, the final RF derived from ($p3, d1$) is ($p3, \{l_1, l_2\})-(p3, d1)-(d1, \{l_2, l_3\}$). After fragment enumeration, both FF and RF procedures can remove some fragments by applying \textcolor{black}{dominance} rules that consider time windows and cost; for more details we refer to \citet{Rist2021}. Finally, it is important to note that the constraints on the maximum number of pickups are integrated into the fragment enumeration process so that the resulting sets $S_{FF}$ and $S_{RF}$ are directly applicable to (M)DARP-LPT\@. Since both fragment sets are generated by enumerating all possible fragments for (M)DARP-LPT, using these sets in fragment-based formulations can achieve optimality.


	\subsection{\textcolor{black}{EFF and ERF generation}}\label{exFrag}
	
	\textcolor{black}{\citet{Rist2022} introduce the concept of EFF, which is a combination of an FF with a node arc. Similarly, we define an ERF as a combination of an RF with a node arc. To construct these EFFs and ERFs, we first generate all FFs and RFs as in Section~\ref{tradFrag}. Then, each fragment is extended by appending either a pickup node or the destination depot. The resulting extended fragments are retained if their corresponding route paths have a feasible time schedule.}
	
	\textcolor{black}{After the extension, the route path of the fragment becomes longer. The end node retains the same location information as the newly added node, while its load information remains unchanged, inheriting the load from the original fragment's end node. Specifically, if the extended node is a pickup node rather than the destination depot, we construct an additional route path by further appending the corresponding delivery location of the extended pickup node. The extended fragment is retained only if this constructed path has a feasible schedule. Other processing steps, such as eliminating dominated fragments, remain unchanged. However, specific procedures such as combining with designated fragments for further feasibility checks are omitted, as the current enumeration process is already time-consuming and may largely exceed the actual solving time.}
	
	\textcolor{black}{An additional point of attention is that when we use EFFs or ERFs, the network flow constraints \textcolor{black}{and the lazy constraints} in formulations FFF and PSFF need to 
		be updated\textcolor{black}{, as most node arcs are eliminated and fragments are directly connected to each other}. Specifically, in FFF we replace constraints~(\ref{FragmentArc1}) and~(\ref{ArcFragment1}) with the revised constraint~\eqref{ex_FFF_flow}, while in PSFF we update constraints~(\ref{flow_pickup}) and~(\ref{flow_delivery})  to~(\ref{ex_PSFF_flow}). \textcolor{black}{The lazy constraints~(\ref{infeasiblePath}) and (\ref{infeasiblecycle}) are also adapted accordingly.}
		\begin{align}
			&&\sum_{f\in F_h^-}x_{fv} &=\sum_{f\in F_h^+}x_{fv} &&  h\in P_N,  v\in V \label{ex_FFF_flow} \\
			&& \sum_{f \in F^-_h} X_{f,l+L_f} &= \sum_{f \in F^+_h} X_{fl} &&   h \in P_N, l \in \mathcal{L_P} \label{ex_PSFF_flow}
	\end{align}}
	
	\subsection{\textcolor{black}{Mixed fragment set construction}}\label{fragmentselect}

	\textcolor{black}{In this subsection we introduce a new fragment set denoted as $S_{MF}$, which consists of a mixture of RFs and ERFs. This set is derived from the original full fragment set $S_{FF}$ and can be a substitute for $S_{FF}$ and $S_{ERF}$. The primary motivation for constructing such a mixed set is to reduce the overall number of fragments while preserving high-quality fragments, thereby accelerating the computation of fragment-based formulations. This new fragment set is motivated by the observation that both fragment length and count affect the quality and efficiency of fragment-based formulations. While longer fragments (e.g., FFs vs.\ RFs, or ERFs vs.\ RFs) strengthen the formulation, their corresponding set is usually larger and potentially slows down the computation, as illustrated in Section~\ref{numericaldiscussion}.}
	
	\textcolor{black}{We observe that many ERFs in $S_{ERF}$ never appear in any EFF in $S_{EFF}$; in particular, some ERFs only combine with specific others to form EFFs that are ultimately dominated. Therefore, rather than exhaustively enumerating all \textcolor{black}{ERFs}, we propose deriving ERFs directly from $S_{EFF}$ or $S_{FF}$. Ultimately, we choose to derive ERFs from $S_{FF}$ for two main reasons. First, enumerating $S_{FF}$ requires less computational effort than $S_{EFF}$. Second, even with reduced redundancy, the number of ERFs generated from $S_{EFF}$ may still be large.	Considering these points, we decide to generate fragments from $S_{FF}$. This, however, results in a mix of ERFs and RFs, rather than pure ERFs. Nevertheless, this approach significantly reduces the total number of fragments and thereby ensures computational efficiency, as demonstrated in Section~\ref{numericaldiscussion}. The generation process and the reason why mixed fragment types arise are further explained below.}
	
	\textcolor{black}{To construct $S_{MF}$, we decompose each FF by identifying pickup locations whose predecessors are delivery locations. If a FF contains a number $w$ of such pickup locations, it is split into $w+1$ pieces. The first $w$ pieces are used to generate ERFs, while the final piece forms an RF\@. For example, consider a fragment with route path $(p_1, p_2, d_2, p_3, d_3, p_4, d_4, d_1)$. This is divided into the following three parts: $(p_1, p_2, d_2, p_3)$, $(p_3, d_3, p_4)$, $(p_4, d_4, d_1)$. These segments are transformed into ERFs and RFs using the method described in Section~\ref{tradFrag}. During this transformation, it is necessary to incorporate load information at the pickup locations. For instance, the locations $p_3$ and $p_4$ are transformed into nodes $(p_3, \{l_1\})$ and $(p_4, \{l_1\})$, respectively. In addition, some FFs cannot be divided and are thus retained (e.g., $(p_1, d_1)$). Therefore, $S_{MF}$ actually includes a mix of ERFs, RFs, and FFs. Since many fragments are RFs or FFs rather than strictly ERFs or EFFs, the fragment set $S_{MF}$ maintains a relatively low overall fragment count.}

	\textcolor{black}{In the resulting network, compared with the full fragment-based network, connections at pickup and delivery nodes with empty loads remain unchanged: fragments connect to node arcs, and node arcs connect to fragments. For pickup nodes with non-empty loads, fragments are directly connected to other fragments. Notably, there are no delivery nodes associated with non-empty loads. As a result, the network flow constraints for FFF and PSFF at nodes with empty loads remain the same as in Section~\ref{FFFormulation} and Section~\ref{PSFFormulation}, respectively, while the constraints at pickup nodes with non-empty loads are updated as described in Section~\ref{exFrag}.}
	
	\textcolor{black}{It is worth noting that if all FFs are decomposed into RFs (instead of ERFs) and node arcs, the resulting RF set is actually a subset of $S_{RF}$. This is because $S_{RF}$ includes all RFs that can be generated from FFs (with the ``\textit{ppppdddd}'' structure) in all possible ways. However, not all RFs in $S_{RF}$ appear in any FF from $S_{FF}$; some can only form dominated FFs when combined with others.}

	\section{Computational results}\label{numericaldiscussion}
	
	\subsection{Data generation and implementation details}\label{IG}
	
	To assess computational efficiency of different formulations and fragment sets, we utilize benchmark instances from \citet{doi:10.1287/opre.1060.0283}, which are widely adopted in studies such as \citet{Ropke2007, Gschwind2015, Rist2021}. These instances comprise two types, A and B, with type A having a tighter time window.
	In type A, customer demand is unit with a vehicle capacity of three, while type B involves demands from one to six customers with a vehicle capacity of six.
	\textcolor{black}{In addition, we utilize benchmark instances from the SDARP problem \citep{Riedler2018,Rist2022}, which feature tighter time windows  (typically three to six times the direct travel time for each customer) compared to those of \citet{doi:10.1287/opre.1060.0283}. There are two main reasons for using this second dataset. First, the tighter time window constraints are more representative of our application scenarios. Second, we aim to evaluate the performance of $S_{EFF}$ for DARP-LPT, as it was shown to perform well for SDARP in \citet{Rist2022}.}
	
	For DARP-LPT, we \textcolor{black}{adapt the aforementioned instances by modifying the number $|V|$ of vehicles in each instance.} Given that DARP-LPT incorporates an additional parameter $L$ for the maximal number of pickups per trip, ensuring coverage for $n$ customers requires a minimum of $\lceil n/L \rceil$ vehicles. Here, we set $|V|$ as $\lceil n/(L-1) \rceil$, which is slightly higher than the minimum required.
	
	For MDARP-LPT, which assumes different drivers have distinct origins and destinations, we generate driver information based on \textcolor{black}{the aforementioned} instances. Specifically, \textcolor{black}{the first $\lceil n/3 \rceil$} customers are designated as drivers, whose pickup and delivery nodes becom\textcolor{black}{e} the origins and destinations, while the remaining customers are passengers. The value $n$ represents the number of customers in each \textcolor{black}{original} instance. For drivers' time windows, we set the earliest departure to 0, maintaining latest arrival times. Unit customer demand and capacity 4 are chosen. 
	
	Table~\ref{table:design} summarizes our experimental design. 
	
	\begin{table}[h]
		\caption{Experimental design}
		\footnotesize
		\centering
		\begin{threeparttable}[b]
			\begin{tabular}{@{}cccccc@{}}
				\toprule
				Problem      & $L$       & $|V|$                         & \textcolor{black}{Cus}                         & Formulation                  & Fragment set                                 \\ \midrule
				DARP-LPT     & $4,6$     & $\lceil n/(L-1) \rceil$       & $n$                         & ABF, FFF\textcolor{black}{, PSFF, PBF}         & $S_{FF}$, $S_{RF}$, \textcolor{black}{$S_{EFF}$, $S_{ERF}$,} $S_{MF}$ \\\midrule
				MDARP-LPT    & $4$       & $\lceil n/3 \rceil$           & $\lfloor 2n/3 \rfloor$      & \textcolor{black}{PSFF, PBF}                     & $S_{FF}$, $S_{RF}$\textcolor{black}{, $S_{MF}$ }                  \\ \bottomrule
			\end{tabular}
			
			\begin{tablenotes}   
				\scriptsize    
				\item[1] \textcolor{black}{Cus = customer number}; Value $n$ is the number of customers in the original instance.
			\end{tablenotes}
		\end{threeparttable}%
		\label{table:design}
	\end{table}
	
	All our algorithms are implemented using the Python programming language. All computational experiments are run on a system with an Intel Core i7-7820HQ processor with 2.90 GHz CPU speed and 32 GB of RAM under a Windows 10 64-bit OS\@. All linear formulations are solved with the commercial solver Gurobi 9.0.3 with a single thread; all other Gurobi parameters are set to their default values. 
	The generation of fragments is implemented using the Python programming language. For DARP-LPT, fragment generation occurs in a single-processing environment, whereas for the MDARP-LPT, fragment generation is accomplished via multi-processing.
	
	
	Since this study involves two problems (DARP-LPT and MDARP-LPT), \textcolor{black}{four different formulations, and five types of fragment sets}, we need to test multiple combinations. Therefore, from the benchmark instances \textcolor{black}{provided by \citet{doi:10.1287/opre.1060.0283}}, we only choose a subset, but we ensure a variety of sizes as indicated by the instance names in all following tables. \textcolor{black}{As the instances from \citet{Riedler2018} have tighter time windows, which adheres more closely to the carpooling and crowdshipping scenarios for the problem addressed in this study, we retain all their instances.}
	
	\subsection{Discussion of the results}
	
	In our evaluation of computational performance, we incorporate several performance indicators: time, OBJ, LB, and gap. Time (s) indicates the total time, including the preprocessing time, network time, and CPU time for solver execution. Here, preprocessing time refers to the duration spent on time window tightening and arc elimination, as detailed by \citet{doi:10.1287/opre.1060.0283}. Network time refers to the time required for fragment generation and the construction of the fragment-based network (in FFF and PSFF), or for path generation (in PBF)\@. OBJ and LB are the best-found integer solution value and the best lower bound within a 30-minute timeframe, respectively, and gap represents the relative percentage difference between OBJ and LB\@. When comparing different fragment sets, we also introduce two additional indicators: fragment count and network time. Fragment count is the total number of fragments in the fragment set.

	\begin{table}[h]
		\caption{\textcolor{black}{Computational performance of different formulations for DARP-LPT (30-minute limit)}}
		\scriptsize  
		\centering
		\setlength{\tabcolsep}{0.5mm}  
		\renewcommand{\arraystretch}{0.95}  
		\resizebox{\textwidth}{!}{  
			\textcolor{black}{			\begin{threeparttable}
					\begin{tabular}{@{}lcccc|cccc|cccc|cccccc@{}}
						\toprule
						& \multicolumn{4}{c|}{ABF} 
						& \multicolumn{4}{c|}{FFF ($S_{FF}$)} 
						& \multicolumn{4}{c|}{PSFF ($S_{FF}$)} 
						& \multicolumn{6}{c}{PBF} \\
						\cmidrule(lr){2-5} \cmidrule(lr){6-9} \cmidrule(lr){10-13} \cmidrule(l){14-19}
						Name & Time & OBJ & LB & Gap 
						& Time & OBJ & LB & Gap 
						& Time & OBJ & LB & Gap 
						& Net & CPU & Time & OBJ & LB & Gap \\
						\midrule
						a3-18 & 24.3 & 302.5 & 302.5 & 0.00\%
						& 0.2 & 302.5 & 302.5 & 0.00\%
						& 0.1 & 302.5 & 302.5 & 0.00\%
						& 0.4 & 0.0 & 0.5 & 302.5 & 302.5 & 0.00\% \\
						a3-30 & 1800.0 & 496.7 & 451.2 & 9.16\%
						& 1800.0 & 494.8 & 491.1 & 0.74\%
						& 1.2 & 494.8 & 494.8 & 0.00\%
						& 2.5 & 0.3 & 3.2 & 494.8 & 494.8 & 0.00\% \\
						a3-36 & 1800.0 & 574.4 & 541.2 & 5.78\%
						& 1800.0 & 574.4 & 565.0 & 1.63\%
						& 0.4 & 574.4 & 574.4 & 0.00\%
						& 6.6 & 1.3 & 9.1 & 574.4 & 574.4 & 0.00\% \\
						a4-48 & 1800.0 & 712.0 & 578.3 & 18.78\%
						& 1800.0 & 697.9 & 687.4 & 1.50\%
						& 0.7 & 697.9 & 697.9 & 0.00\%
						& 26.0 & 3.0 & 33.2 & 697.9 & 697.9 & 0.00\% \\
						a6-60 & 1800.0 & 920.2 & 635.4 & 30.95\%
						& 1800.0 & 864.9 & 843.4 & 2.48\%
						& 2.7 & 864.9 & 864.9 & 0.00\%
						& 5.2 & 1.0 & 7.1 & 864.9 & 864.9 & 0.00\% \\
						a7-70 & 1800.0 & 1284.3 & 664.6 & 48.25\%
						& 1800.0 & 986.6 & 956.5 & 3.05\%
						& 8.3 & 984.1 & 984.0 & 0.00\%
						& 9.9 & 1.8 & 13.4 & 984.1 & 984.0 & 0.00\% \\
						a8-80 & \multicolumn{4}{c|}{No}
						& 1800.0 & 1071.1 & 1019.3 & 4.84\%
						& 10.0 & 1058.7 & 1058.7 & 0.00\%
						& 25.9 & 5.6 & 35.7 & 1058.7 & 1058.7 & 0.00\% \\
						a8-96 & \multicolumn{4}{c|}{No}
						& 1800.0 & 1353.6 & 1302.7 & 3.76\%
						& 13.3 & 1342.2 & 1342.2 & 0.00\%
						& \multicolumn{6}{c}{No} \\
						\midrule
						Avg-A & 1504.1 &    &    & 18.82\%
						& 1512.9 &    &    & 2.25\%
						& 4.59 &    &    & 0.00\%
						& 14.6 &    &    &    &    & 0.00\% \\
						\midrule
						b3-36 & 1800.0 & 576.5 & 548.4 & 4.88\%
						& 1800.0 & 576.5 & 567.7 & 1.53\%
						& 0.4 & 576.5 & 576.5 & 0.00\%
						& 59.4 & 12.4 & 83.0 & 576.5 & 576.5 & 0.00\% \\
						b4-40 & 1800.0 & 637.5 & 612.5 & 3.92\%
						& 1800.0 & 635.2 & 631.1 & 0.64\%
						& 1.0 & 635.2 & 635.2 & 0.00\%
						& 121.0 & 35.2 & 174.6 & 635.2 & 635.2 & 0.00\% \\
						b5-50 & 1800.0 & 796.5 & 659.3 & 17.22\%
						& 1800.0 & 789.1 & 768.3 & 2.63\%
						& 3.1 & 789.1 & 789.1 & 0.00\%
						& 239.9 & 95.8 & 371.5 & 789.1 & 789.1 & 0.00\% \\
						b6-60 & 1800.0 & 937.9 & 749.9 & 20.05\%
						& 1800.0 & 931.7 & 900.7 & 3.33\%
						& 4.9 & 927.1 & 927.1 & 0.00\%
						& 59.9 & 25.9 & 95.8 & 927.1 & 927.1 & 0.00\% \\
						b7-70 & 1800.0 & 1162.9 & 799.9 & 31.21\%
						& 1800.0 & 950.4 & 920.7 & 3.13\%
						& 7.9 & 944.7 & 944.7 & 0.00\%
						& 105.1 & 33.9 & 158.0 & 944.7 & 944.7 & 0.00\% \\
						b8-80 & 1800.0 & 1741.53 & 889.69 & 0.49\% 
						& 1800.0 & 1104.8 & 1072.0 & 2.97\% 
						& 3.6 & 1103.2 & 1103.2 & 0.00\%
						& \multicolumn{6}{c}{No} \\
						b8-96 & \multicolumn{4}{c|}{No} 
						& 1800.0 & 1322.6 & 1260.9 & 4.67\% 
						& 7.4 & 1294.4 & 1294.4 & 0.00\%
						& \multicolumn{6}{c}{No} \\
						\midrule
						Avg-B & 1800.0 &    &    & 12.96\%
						& 1800.0 &    &    & 2.70\%
						& 4.04 &    &    & 0.00\%
						& 176.6 &    &    &    &    & 0.00\% \\
						\bottomrule
					\end{tabular}
					\begin{tablenotes}
						\item[1] Net: network time; CPU: CPU time for solver. The rows labeled Avg-A and Avg-B represent the average metric values calculated across all instances of type A and type B, respectively. ``No'' indicates that no feasible solution was obtained within the 30-minute time limit. 
					\end{tablenotes}
			\end{threeparttable}}
		}  
		\label{table:darpextend1}
	\end{table}

	\textcolor{black}{Table~\ref{table:darpextend1} compares the four formulations ABF, FFF, PSFF, and PBF for DARP-LPT for a runtime limit of 30 minutes, using the instances of \citet{doi:10.1287/opre.1060.0283} with the parameters $|V|=\lceil n/3 \rceil$ and $L=4$. Both FFF and PSFF utilize $S_{FF}$ as the input fragment set. The table indicates that, as observed from the time and gap indicators, FFF and PSFF largely outperform ABF, with PSFF exhibiting superior performance compared to FFF\@. Even with $L = 4$, PBF performs worse than PSFF in terms of the total time. For larger instances (e.g., a8-96 and b8-96), PBF does not have sufficient time to generate all the required paths.}

	\begin{table}[h]
		\caption{\textcolor{black}{Average computational performance of different fragment sets for DARP-LPT and MDARP-LPT (30-minute limit)}}
		\centering
		\scriptsize
		\setlength{\tabcolsep}{0.4 mm}
		\renewcommand{\arraystretch}{1.0}
		\textcolor{black}{		\begin{threeparttable}
				\begin{tabular}{lcc|ccc|ccc|ccc|ccc|ccc}
					\toprule
					\multicolumn{3}{c|}{} & \multicolumn{3}{c|}{$S_{RF}$} & \multicolumn{3}{c|}{$S_{FF}$} & \multicolumn{3}{c|}{$S_{MF}$} & \multicolumn{3}{c|}{$S_{EFF}$} & \multicolumn{3}{c}{$S_{ERF}$} \\\midrule
					Problem & $L$ & Type & F & CPU & T & F & CPU & T & F & CPU & T & F & CPU & T & F & CPU & T \\
					\midrule
					\multirow{4}{*}{\makecell[c]{DARP\\-LPT}}  & 4 & A & 599.4 & 0.6 & 1.5 & 713.0 & 1.0 & 4.5 & 700.3 & 1.1 & 3.6 & 15352.3 & 1.5 & 21.5 & 46594.0 & 6.6 & 27.9 \\
					~ & ~ & B & 1231.9 & 1.9 & 3.2 & 1777.7 & 1.7 & 9.4 & 1287.6 & 1.8 & 7.9 & 24920.4 & 2.7 & 54.9 & 103744.7 & 7.2 & 55.6 \\
					~ & 6 & A & 599.4 & 9.9 & 11.2 & 1153.0 & 10.8 & 21.4 & 707.6 & 11.1 & 20.4 & 29132.4 & 19.5 & 69.5 & 46594.0 & 207.7 & 229.9 \\
					~ & ~ & B & 1231.9 & 9.9 & 12.0 & 9905.6 & 20.5 & 164.1 & 1439.0 & 16.3 & 170.3 & \multicolumn{3}{c|}{6/7 solved} & 108323.7 & 46.7 & 104.1 \\
					\midrule
					\multirow{2}{*}{\makecell[c]{MDARP\\-LPT}}  & 4 & A & 252.5 & 19.0 & 35.0 & 252.5 & 20.1 & 39.1 & 264.9 & 19.2 & 36.5 & \multicolumn{3}{c|}{--} & \multicolumn{3}{c}{--} \\
					~ & ~ & B & 2079.6 & 82.4 & 124.4 & 2285.4 & 47.5 & 126.9 & 1708.4 & 71.4 & 154.8 & \multicolumn{3}{c|}{--} & \multicolumn{3}{c}{--} \\
					\midrule
					\multirow{2}{*}{\makecell[c]{DARP\\-LPT}}  & 4 & \multirow{2}{*}{\makecell[c]{Ried\\-ler}} & 1301.5 & 3.4 & 4.8 & 2328.4 & 2.1 & 6.3 & 1702.9 & 2.3 & 6.2 & 17471.0 & 2.6 & 18.0 & 60836.1 & 19.5 & 46.0 \\
					~ & 6 & ~ & 1301.5 & 43.8 & 45.6 & 11845.3 & 49.6 & 83.2 & 1782.6 & 31.0 & 63.0 & 73336.5 & 64.4 & 172.8 & \multicolumn{3}{c}{26/30 solved} \\
					\midrule
					\makecell{MDARP\\-LPT} & 4 &  \makecell{Ried\\-ler} & 454.7 & 17.2 & 31.6 & 633.87 & 14.6 & 31.9 & 547.0 & 14.3 & 32.0 & \multicolumn{3}{c|}{--} & \multicolumn{3}{c}{--} \\
					\bottomrule
				\end{tabular}
				\begin{tablenotes}
					\item[1] Type denotes the instance type: A and B refer to type A and type B instances from \citet{doi:10.1287/trsc.1090.0272}, respectively; Riedler refers to instances from \citet{Riedler2018}.  
					F: fragment count; CPU: CPU time for solver; T: Total time (s); ``--'' denotes that no further testing is performed for this scenario.
				\end{tablenotes}
				\vspace{1mm}
		\end{threeparttable}}
		\label{table:darpextend2}
	\end{table}

	\textcolor{black}{Table~\ref{table:darpextend2} presents a comparative analysis of the five fragment sets $S_{RF}$, $S_{FF}$, $S_{MF}$, $S_{EFF}$, and $S_{ERF}$ for both DARP-LPT and MDARP-LPT\@. All these fragment sets are utilized as input for formulation PSFF\@. The table covers the instances from \citet{doi:10.1287/opre.1060.0283} and \citet{Riedler2018}, with varying $L$. Due to the extensive amount of results, we present only the average metrics (fragment count, CPU time, and total time) for all instances here. The total time mainly comprises network time and CPU time, with all other components contributing less than 1\%. Since both CPU time and total time are reported, we do not report network time separately. Detailed results for all instances are provided in \ref{resultDARPLPT}.}
	
	\textcolor{black}{The table provides several key insights. Our analysis focuses on three main aspects: determining the best fragment set for the current (M)DARP-LPT problem, evaluated by total time; assessing fragment set quality, measured by CPU time alone; and explaining why different fragment sets exhibit varying performance. }
	
	\textcolor{black}{Table~\ref{table:darpextend2} leads to several noteworthy observations. First, when fed into the formulation PSFF, the fragment set $S_{RF}$ consistently outperforms alternative sets in terms of the total time to reach optimality across all problem instances. Second, when CPU time is considered independently (that is, excluding network time) $S_{FF}$ achieves performance comparable to or even surpassing that of $S_{RF}$ for instances with tight time windows and low values of $L$ (e.g., in DARP-LPT and MDARP-LPT, for instances from \citet{Riedler2018} with $L=4$). By contrast, in DARP-LPT, for type-B instances with $L=6$, $S_{RF}$ demonstrates a clear advantage over $S_{FF}$, primarily due to the substantially higher fragment count of $S_{FF}$. Third, $S_{MF}$ proves to be an effective alternative of $S_{FF}$ when the fragment count of $S_{FF}$ becomes prohibitively large. For example, for \citeauthor{Riedler2018}'s instances in DARP-LPT with $L=6$, $S_{MF}$ is the most effective fragment set in terms of CPU time performance. Fourth, both $S_{EFF}$ and $S_{ERF}$ demonstrate inferior performance in terms of the total time to reach optimality, even when considering CPU time alone. This is primarily due to their excessive fragment counts. Analogously to the comparison between $S_{FF}$ and $S_{RF}$, $S_{EFF}$ achieves better performance than $S_{ERF}$ when $L$ is small and the instances have tight time windows and ride times. Given that $S_{ERF}$ and $S_{EFF}$ demonstrate significantly inferior performance compared to other fragment sets in DARP-LPT, we do not further evaluate them for MDARP-LPT.}
	
	\textcolor{black}{The performance differences observed across fragment sets can be largely attributed to two factors: fragment count and the quality of the resulting lower bound. The comparison between $S_{FF}$ and $S_{RF}$ is particularly illustrative of this underlying logic. When $L$ is lower, or when time windows or maximum ride time are tight, the fragment counts for $S_{FF}$ and $S_{RF}$ are similar or at least of the same magnitude. Conversely, when $L$ is higher and both time windows and maximum ride time are more flexible, the fragment count for $S_{FF}$ is substantially higher than that for $S_{RF}$. This is important because employing a large fragment set as input can slow down the computation process. Additionally, the fragments from $S_{FF}$ are longer than those from $S_{RF}$, in a similar way as paths are longer than individual arcs. This similarity helps to explain why $S_{FF}$ generally achieves better lower bounds than $S_{RF}$, because path-based formulations for routing problems typically yield superior lower bounds compared to arc-based formulations \citep{G-2024-36}. Additionally, when transitioning to $S_{MF}$, the set effectively shifts fragments from $S_{FF}$ to ERFs \textcolor{black}{and RFs}. As a result, the average fragment length in $S_{MF}$ lies between that of $S_{FF}$ and $S_{RF}$, while the fragment count remains of the same order of magnitude as that of $S_{RF}$. This balance explains why $S_{MF}$ achieves superior CPU time performance in certain scenarios.}
	
	\textcolor{black}{Given the relative simplicity of the current problem and the efficiency of the PSFF formulation, the benefits of higher-quality fragments are outweighed by the overhead of fragment enumeration. However, under the slower FFF formulation, the advantage of $S_{FF}$ over $S_{RF}$ becomes more pronounced. For example, in DARP-LPT with $L=4$, using the FFF formulation with a 30-minute time limit (but excluding the instance b8-96, where $S_{RF}$ fails to produce a feasible solution), $S_{FF}$ achieves average optimality gaps of 2.25\% (type A) and 2.37\% (type B), compared to 2.43\% and 2.53\% for $S_{RF}$, respectively (see \ref{resultDARPLPTFFF}). Moreover, when the problem becomes more realistic by introducing additional constraints, the fragment enumeration time tends to remain stable (as the number of feasible paths becomes even smaller), whereas the solution time grows exponentially. This underscores the increasing value of employing higher-quality fragment sets as problem difficulty increases. Additionally, even without considering more complex settings but simply increasing the instance size, $S_{FF}$ outperforms $S_{RF}$ in terms of total time; this can be seen, for example, in DARP-LPT with \citeauthor{Riedler2018}'s largest instance $60N\_5K\_C$ and $L = 6$ (where both $S_{MF}$ and $S_{FF}$ even incur higher network construction time). Although the average metrics may not fully capture the importance of fragment set quality, the results on larger instances indicate that more difficult cases favor good-quality fragment sets. One further thing to mention is that some shared mobility studies \citep{Zhang2022,Su2024} use EFFs or FFs for DARP or similar problems. While the enumeration time is quite high in our study, it is significantly lower in theirs. This is partly because their instances feature much tighter time windows, and partly because we invest additional time in fragment elimination to ensure the best possible fragment set.}

	\section{Summary and conclusions}\label{conclusion}
	In this paper we have introduced two variants of the dial-a-ride problem, specifically MDARP-LPT and DARP-LPT\@. MDARP-LPT addresses scenarios where vehicles have unique origins and destinations, while DARP-LPT assumes that all vehicles begin and end their routes at a fixed depot. These problem variants are motivated by real-world carpooling and crowdshipping scenarios, where each trip can pick up at most a certain number $L$ of customers. (M)DARP-LPT has been proven to be NP-hard through a reduction from the capacitated vehicle routing problem.
	
	To efficiently solve both MDARP-LPT and DARP-LPT, we have applied a unified fragment-based method, where a fragment is a partial path. Our main contributions entail the extensions of a fragment-based formulation, the creation of improved fragment sets, and the insights gained from comparing fragment sets. More specifically, we model (M)DARP-LPT using two fragment-based formulations, FFF \textcolor{black}{and PSFF}, both of which are extended from the DARP fragment-based framework proposed by \citet{Rist2021}. \textcolor{black}{For FFF the extension resides in the transition} from fragment-based to fragment–vehicle–based decision variables\textcolor{black}{, whereas PSFF retains the original variables but augments the network with a pickup-count dimension, resulting in a pickup-space network. We find that PSFF achieves better results than FFF\@. We also compare several fragment sets used as inputs to the formulations. The full fragment set $S_{FF}$ and the restricted fragment set $S_{RF}$ are generated following the methods proposed by \citet{Alyasiry2019} and \citet{Rist2021}, respectively. The extended full fragment set $S_{EFF}$ and the extended restricted fragment set $S_{ERF}$ are constructed based on the approach of \citet{Rist2022}. In addition, we develop the new mixed fragment set $S_{MF}$ by decomposing full fragments from $S_{FF}$ into (extended) restricted fragments. All these fragment sets can be used as inputs to fragment-based formulations to reach optimality. Our computational results show that, for both DARP-LPT and MDARP-LPT, $S_{RF}$ outperforms $S_{FF}$ in total solution time when used with the PSFF formulation. However, in terms of fragment set quality, measured solely by CPU time, $S_{FF}$ remains superior in instances with tight time windows and a limited number of pickups (which is also close to realistic application scenarios). In DARP-LPT, $S_{MF}$ serves as a high-quality alternative to $S_{FF}$, especially when the size of $S_{FF}$ becomes large.}
	
	\textcolor{black}{The incorporation of a time dimension into the formulations FFF or PSFF may be a valuable direction for future work. For example, PSFF can naturally be extended to a pickup-time-space formulation to incorporate both time and pickup dimensions.} However, since vehicle indices and pickup count are already used as dimensions, and PSFF performs well under the current setup, we have not pursued this extension in this paper.
	
	\bibliographystyle{model5names}
	\biboptions{authoryear}
	\bibliography{library2}
	
	\clearpage
	\appendix
	
	\setcounter{figure}{0}
	\renewcommand{\thefigure}{A.\arabic{figure}}
	
	\section{Additional explanation for fragment-based network}\label{AEFBN}

	Decomposing feasible DARP routes into fragments yields a significantly smaller number of fragments compared to DARP routes. Conversely, if we combine these fragments, we can obtain all feasible DARP routes, albeit with the possibility of some redundant routes. Fig.~\ref{fig2} illustrates how feasible DARP routes can be gathered in a transitional network (as an intermediate step towards the fragment-based network). Fig.~\ref{fig1} presents several DARP routes and their corresponding fragments. The network in Fig.~\ref{fig2} will use these fragments and allow to enumerate all the DARP routes shown in Fig.~\ref{fig1}. Identical fragments across different routes in Fig.~\ref{fig1} can be consolidated into a single fragment while maintaining the connections to other fragments. Subsequently, the fragment-based network of Fig.~\ref{fig3} is derived from the network of Fig.~\ref{fig2}. When different fragments share the same start and/or end nodes, then those nodes can also be merged. For instance, two fragments with the same start and end nodes but different route paths $(p1,d1)$ and $(p1,p4,d4,d1)$ become two route paths originating from node $p1$ and terminating at node $d1$. Such consolidations help reduce connections (node arcs). Node arcs with the same start node and end node can also be merged; the three connections from $(p1,d1)$ to $(p2,d2)$, from $(p1,d1)$ to $(p2,p3,d2,d3)$, and from $(p1,p4,d4,d1)$ to $(p2,d2)$ in Fig.~\ref{fig2}, for instance, can be merged into one node arc $(d1,p2)$ in Fig.~\ref{fig3}.
	
	\begin{figure}[h]
		\centering
		\resizebox{1.0\textwidth}{!}{
			\begin{tikzpicture}
				\node[draw,ellipse] (0+) at (-5.5,0) {$O^+$};
				\node[draw, ellipse] (p3) at (-0.5,2) {$p3$};
				\node[draw, ellipse] (d2p3) at (3.5,2) {($d2$,\{$l_3$\})};
				\node[draw, ellipse] (p1p3) at (6.5,2) {($p1$,\{$l_3$\})};
				\node[draw, ellipse] (d3) at (10,2) {$d3$};
				\node[draw, ellipse] (0-) at (11.5,0) {$O^-$};
				
				\draw[line width=0.75pt, ->,dashed,>=latex] (0+) -> (p3);
				\draw[line width=0.75pt, ->,dashed,>=latex] (d3) ->  (0-);
				\draw[line width=0.75pt, ->,dashed,>=latex] (d2p3) -> (p1p3);
				
				\draw[line width=0.75pt, ->,>=latex] (p3) -- node[above] {($p3$,$p2$,$d2$)} (d2p3);
				\draw[line width=0.75pt, ->,>=latex] (p1p3) -- node[above] {($p1$,$d1$,$d3$)} (d3);

				\node[draw, ellipse] (1p1) at (-4,0) {$p1$};
				\node[draw, ellipse] (1d1) at (-2,0) {$d1$};
				\node[draw, ellipse] (1p2) at (1,0) {$p2$};
				\node[draw, ellipse] (1d2) at (4,0) {$d2$};
				\node[draw, ellipse] (1p3) at (6,0) {$p3$};
				\node[draw, ellipse] (1d3) at (9,0) {$d3$};
				
				\draw[line width=0.75pt, ->,dashed,>=latex] (0+) -> (1p1);
				\draw[line width=0.75pt, ->,dashed,>=latex] (1d3) ->  (0-);
				\draw[line width=0.75pt, ->,dashed,>=latex] (1d1) -> (1p2);
				\draw[line width=0.75pt, ->,dashed,>=latex] (1d2) -> (1p3);
				
				\draw[line width=0.75pt, ->,>=latex] (1p1) -- node[above] {($p1$,$d1$)} (1d1);
				\draw[line width=0.75pt, ->,>=latex] (1p2) -- node[above] {($p2$,$d2$)} (1d2);
				\draw[line width=0.75pt, ->,>=latex] (1p3) -- node[above] {($p3$,$d3$)} (1d3);
				
				\draw[line width=0.75pt, ->,dashed,>=latex] (1d2) to[out=-15, in=-165] (0-);
				
				\node[draw, ellipse] (4p1) at (-4,-2) {$p1$};
				\node[draw, ellipse] (4d1) at (-0.5,-2) {$d1$};
				\node[draw, ellipse] (4p2) at (2,-2) {$p2$};
				\node[draw, ellipse] (4d3) at (7,-2) {$d3$};
				
				\draw[line width=0.75pt, ->,dashed,>=latex] (0+) -> (4p1);
				\draw[line width=0.75pt, ->,dashed,>=latex] (4d3) ->  (0-);
				\draw[line width=0.75pt, ->,dashed,>=latex] (4d1) -> (1p3);
				\draw[line width=0.75pt, ->,dashed,>=latex] (4d1) -> (1p2);
				\draw[line width=0.75pt, ->,dashed,>=latex] (1d1) -> (4p2);
				
				\draw[line width=0.75pt, ->,>=latex] (4p1) -- node[above] {($p1$,$p4$,$d4$,$d1$)} (4d1);
				\draw[line width=0.75pt, ->,>=latex] (4p2) -- node[above] {($p2$,$p3$,$d2$,$d3$)}(4d3);
				
		\end{tikzpicture}}
		\caption{A transitional network for DARP by merging all identical fragments from Fig.~\ref{fig1}}\label{fig2}
	\end{figure}

	\section{Full computational results of PSFF for DARP-LPT and MDARP-LPT}\label{resultDARPLPT}
	
	\setcounter{table}{0}
	
	The tables below provide supplementary information an the results for PSFF, accompanying Table~\ref{table:darpextend2}. As nearly all instances achieve optimality across different fragment set inputs, we report only one objective value (OBJ) in the first column and keep four indicators: fragment count, network time, CPU time, and the total time. The test instances comprises: (1) eight Type A and seven Type B instances from \citet{doi:10.1287/opre.1060.0283}, and (2) 30 instances from \citet{Riedler2018}.
	
	Tables~\ref{table:darp311}--\ref{table:darp313} present the computational performance of PSFF with different fragment sets on the instances from \citet{doi:10.1287/opre.1060.0283}. Specifically, Tables~\ref{table:darp311} and \ref{table:darp312} correspond to DARP-LPT, while Table~\ref{table:darp313} focuses on MDARP-LPT.
	
	Similarly, Tables~\ref{table:darp321}--\ref{table:darp323} evaluate PSFF's performance on the instances from \citet{Riedler2018}, with Tables~\ref{table:darp321} and \ref{table:darp322} dedicated to DARP-LPT and Table~\ref{table:darp323} to MDARP-LPT.

	\begin{table}[H]
		\hspace{0cm}
		\makebox[\textwidth][c]{%
			\begin{threeparttable}
				\caption{Computational performance of PSFF with different fragment sets for DARP-LPT within 30 minutes (\citeauthor{doi:10.1287/opre.1060.0283}'s instances, $L=4$)}
				\footnotesize
				\setlength{\tabcolsep}{0.2mm}
				\begin{tabular}{@{}cccccc|cccc|cccc|cccc|cccc@{}}
					\toprule
					&        & \multicolumn{4}{c|}{$S_{RF}$} & \multicolumn{4}{c|}{$S_{FF}$} & \multicolumn{4}{c|}{$S_{MF}$} & \multicolumn{4}{c|}{$S_{EFF}$} & \multicolumn{4}{c}{$S_{ERF}$}   \\ \midrule
					Name  & OBJ    & F       & Net   & CPU  & Time & F       & Net   & CPU  & Time & F       & Net   & CPU  & Time & F       & Net   & CPU  & Time  & F        & Net   & CPU  & Time  \\ \midrule
					a3-18 & 302.5  & 73      & 0.1   & 0.0  & 0.1  & 64      & 0.1   & 0.1  & 0.2  & 71      & 0.1   & 0.0  & 0.2  & 455     & 0.2   & 0.0  & 0.3   & 1310     & 0.4   & 0.0  & 0.5   \\
					a3-30 & 494.8  & 93      & 0.1   & 0.1  & 0.2  & 89      & 0.2   & 0.1  & 0.4  & 95      & 0.1   & 0.1  & 0.3  & 1202    & 0.8   & 0.1  & 0.9   & 2902     & 0.8   & 0.1  & 1.1   \\
					a3-36 & 574.4  & 121     & 0.1   & 0.1  & 0.3  & 112     & 0.1   & 0.1  & 0.4  & 122     & 0.2   & 0.2  & 0.4  & 1730    & 1.1   & 0.1  & 1.5   & 4261     & 1.1   & 0.2  & 1.7   \\
					a4-48 & 697.9  & 318     & 0.2   & 0.3  & 0.7  & 314     & 0.9   & 0.1  & 1.2  & 338     & 0.8   & 0.1  & 1.3  & 4943    & 5.6   & 0.2  & 6.1   & 14468    & 5.5   & 0.4  & 7.2   \\
					a6-60 & 864.9  & 601     & 0.3   & 0.4  & 1.1  & 655     & 3.4   & 0.6  & 4.6  & 670     & 1.7   & 0.4  & 2.5  & 14542   & 12.5  & 1.5  & 15.0  & 48423    & 17.2  & 1.9  & 23.8  \\
					a7-70 & 984.1  & 1036    & 0.7   & 0.8  & 2.1  & 1306    & 4.0   & 1.9  & 6.8  & 1211    & 2.4   & 1.0  & 3.8  & 19146   & 22.5  & 2.1  & 26.5  & 57082    & 21.1  & 14.4 & 41.4  \\
					a8-80 & 1058.7 & 1088    & 0.6   & 1.1  & 2.4  & 1480    & 7.1   & 0.9  & 8.7  & 1386    & 4.5   & 0.9  & 5.8  & 33590   & 47.3  & 2.6  & 52.6  & 89453    & 30.3  & 15.2 & 56.0  \\
					a8-96 & 1342.2 & 1465    & 1.37  & 2.2  & 5.0  & 1684    & 7.9   & 4.3  & 13.3 & 1709    & 7.2   & 6.3  & 14.3 & 47210   & 59.5  & 5.5  & 68.8  & 154853   & 55.0  & 20.3 & 91.6  \\ \midrule
					Avg-A &        & 599.4   &       & 0.6  & 1.5  & 713.0   &       & 1.0  & 4.5  & 700.3   &       & 1.1  & 3.6  & 15352.3 &       & 1.5  & 21.5  & 46594.0  &       & 6.6  & 27.9  \\ \midrule
					b3-36 & 576.5  & 169     & 0.1   & 0.1  & 0.3  & 166     & 0.2   & 0.2  & 0.5  & 168     & 0.2   & 0.1  & 0.4  & 1473    & 1.4   & 0.1  & 1.6   & 3834     & 1.2   & 0.1  & 1.6   \\
					b4-40 & 635.2  & 311     & 0.2   & 0.1  & 0.4  & 212     & 1.3   & 0.1  & 1.6  & 233     & 0.5   & 0.1  & 0.8  & 3572    & 3.9   & 0.2  & 4.3   & 14683    & 4.8   & 0.3  & 6.2   \\
					b5-50 & 789.1  & 518     & 0.3   & 0.7  & 1.2  & 804     & 1.6   & 0.4  & 2.5  & 649     & 1.4   & 0.6  & 2.3  & 9808    & 7.4   & 0.7  & 8.7   & 19161    & 6.3   & 1.0  & 9.3   \\
					b6-60 & 927.1  & 940     & 0.8   & 2.3  & 3.7  & 1231    & 2.8   & 1.6  & 4.8  & 1030    & 1.9   & 1.1  & 3.4  & 12290   & 15.1  & 2.1  & 18.6  & 28817    & 9.1   & 2.1  & 13.4  \\
					b7-70 & 944.7  & 960     & 0.8   & 4.6  & 5.9  & 1678    & 4.5   & 3.5  & 8.7  & 1188    & 3.3   & 2.4  & 6.0  & 18758   & 24.0  & 2.6  & 28.5  & 52404    & 21.2  & 3.7  & 30.0  \\
					b8-80 & 1103.2 & 1195    & 0.8   & 2.3  & 3.6  & 1813    & 5.2   & 1.5  & 7.2  & 1429    & 4.4   & 3.0  & 7.9  & 26484   & 43.4  & 3.0  & 49.3  & 71693    & 27.0  & 3.4  & 37.0  \\
					b8-96 & 1294.4 & 4530    & 3.1   & 3.1  & 7.4  & 6540    & 34.9  & 4.4  & 40.8 & 4316    & 28.5  & 5.3  & 34.9 & 102058  & 256.3 & 10.0 & 273.5 & 535621   & 204.4 & 39.5 & 291.9 \\ \midrule
					Avg-B &        & 1231.9  &       & 1.9  & 3.2  & 1777.7  &       & 1.7  & 9.4  & 1287.6  &       & 1.8  & 7.9  & 24920.4 &       & 2.7  & 54.9  & 103744.7 &       & 7.2  & 55.6  \\ \bottomrule
				\end{tabular}
				\label{table:darp311}
				\begin{tablenotes} \scriptsize
					\item[1] F: fragment count; Net: network time, the time required for fragment generation and fragment-based network construction; CPU: CPU time for solver; Time~(s) represents the total time, including preprocessing time, network time, and CPU time for solver. The rows labeled Avg-A and Avg-B represent the average metric values calculated across all instances of type A and type B, respectively.
				\end{tablenotes}
			\end{threeparttable}
		}
	\end{table}

	\begin{table}[H]
		\hspace{0cm}
		\makebox[\textwidth][c]{%
			\begin{threeparttable}
				\caption{Computational performance of PSFF with different fragment sets for DARP-LPT within 30 minutes (\citeauthor{doi:10.1287/opre.1060.0283}'s instances, $L=6$)}
				\footnotesize
				\setlength{\tabcolsep}{0.2 mm}
				\begin{tabular}{@{}cccccc|cccc|cccc|cccc|cccc@{}}
					\toprule
					&        & \multicolumn{4}{c|}{$S_{RF}$} & \multicolumn{4}{c|}{$S_{FF}$} & \multicolumn{4}{c|}{$S_{MF}$} & \multicolumn{4}{c|}{$S_{EFF}$} & \multicolumn{4}{c}{$S_{ERF}$}   \\ \midrule
					Name  & OBJ    & F       & Net  & CPU   & Time & F      & Net   & CPU   & Time   & F      & Net   & CPU  & Time   & F       & Net   & CPU   & Time  & F        & Net   & CPU    & Time   \\ \midrule
					a3-18 & 295.8  & 73      & 0.0  & 0.0   & 0.1  & 64     & 0.5   & 0.2   & 1.0    & 71     & 0.1   & 0.0  & 0.1    & 470     & 0.2   & 0.0   & 0.3   & 1310     & 0.4   & 0.1    & 0.6    \\
					a3-30 & 477.0  & 93      & 0.1  & 0.2   & 0.3  & 89     & 0.6   & 0.9   & 1.8    & 95     & 0.1   & 0.2  & 0.7    & 1250    & 0.7   & 0.3   & 3.1   & 2902     & 0.9   & 0.5    & 1.8    \\
					a3-36 & 549.8  & 121     & 0.1  & 0.1   & 0.3  & 115    & 0.4   & 0.1   & 0.9    & 122    & 0.2   & 0.1  & 0.4    & 1951    & 1.5   & 0.2   & 1.9   & 4261     & 1.4   & 0.2    & 2.3    \\
					a4-48 & 669.4  & 318     & 0.2  & 0.6   & 1.1  & 344    & 1.8   & 0.5   & 3.4    & 339    & 0.9   & 0.5  & 1.6    & 5818    & 7.4   & 1.3   & 9.5   & 14468    & 4.9   & 4.1    & 11.3   \\
					a6-60 & 825.0  & 601     & 0.3  & 0.9   & 1.7  & 827    & 5.2   & 1.1   & 6.8    & 673    & 3.5   & 0.8  & 4.7    & 23568   & 34.0  & 4.4   & 44.4  & 48423    & 14.1  & 19.5   & 41.3   \\
					a7-70 & 925.7  & 1036    & 0.7  & 6.6   & 8.3  & 2157   & 15.5  & 3.1   & 19.5   & 1233   & 13.3  & 2.6  & 16.5   & 38512   & 66.8  & 10.8  & 84.7  & 57082    & 17.2  & 126.5  & 153.1  \\
					a8-80 & 990.8  & 1088    & 0.9  & 8.9   & 11.1 & 3028   & 28.6  & 7.7   & 37.3   & 1390   & 25.5  & 6.7  & 33.0   & 81310   & 123.8 & 24.5  & 157.7 & 89453    & 26.7  & 317.5  & 359.6  \\
					a8-96 & 1262.3 & 1465    & 1.0  & 61.7  & 66.7 & 2600   & 25.8  & 73.2  & 100.1  & 1738   & 26.5  & 78.2 & 105.9  & 80180   & 128.0 & 114.4 & 254.6 & 154853   & 49.2  & 1192.9 & 1269.1 \\ \midrule
					Avg-A &        & 599.4   &      & 9.9   & 11.2 & 1153.0 &       & 10.8  & 21.4   & 707.6  &       & 11.1 & 20.4   & 29132.4 &       & 19.5  & 69.5  & 46594.0  &       & 207.7  & 229.9  \\ \midrule
					b3-36 & 557.9  & 169     & 0.1  & 0.1   & 0.3  & 219    & 1.1   & 0.2   & 1.7    & 172    & 0.3   & 0.1  & 0.5    & 1723    & 2.0   & 0.1   & 2.3   & 3857     & 1.2   & 0.2    & 2.1    \\
					b4-40 & 627.4  & 311     & 0.2  & 0.7   & 1.1  & 238    & 5.0   & 1.0   & 6.3    & 244    & 2.6   & 1.0  & 3.8    & 5033    & 12.6  & 1.2   & 14.5  & 14683    & 4.8   & 1.9    & 8.8    \\
					b5-50 & 751.9  & 518     & 0.4  & 3.4   & 4.3  & 1869   & 8.6   & 1.3   & 10.4   & 653    & 5.5   & 0.9  & 6.6    & 31667   & 48.9  & 4.6   & 57.3  & 19197    & 5.8   & 1.6    & 10.4   \\
					b6-60 & 876.8  & 940     & 0.5  & 3.7   & 5.0  & 3818   & 16.5  & 2.1   & 19.3   & 1079   & 23.6  & 2.1  & 26.2   & 30029   & 73.5  & 10.7  & 89.8  & 28817    & 10.5  & 5.3    & 19.7   \\
					b7-70 & 896.4  & 960     & 1.0  & 2.2   & 4.5  & 6251   & 28.7  & 1.8   & 31.4   & 1252   & 38.3  & 2.1  & 41.1   & 67029   & 149.1 & 7.3   & 163.2 & 52603    & 23.0  & 6.8    & 38.4   \\
					b8-80 & 1053.8 & 1195    & 1.1  & 12.2  & 14.5 & 5896   & 34.9  & 14.2  & 50.1   & 1481   & 39.7  & 10.8 & 51.4   & 78389   & 173.2 & 14.8  & 196.8 & 71693    & 24.9  & 14.5   & 52.4   \\
					b8-96 & 1222.0 & 4530    & 3.8  & 47.1  & 54.1 & 51048  & 902.7 & 122.7 & 1029.7 & 5192   & 963.1 & 97.5 & 1062.4 & \multicolumn{4}{c|}{No}         & 567416   & 212.9 & 296.4  & 597.0  \\ \midrule
					Avg-B &        & 1231.9  &      & 9.9   & 12.0 & 9905.6 &       & 20.5  & 164.1  & 1439.0 &       & 16.3 & 170.3  & \multicolumn{4}{c|}{6/7 solved}   & 108323.7 &       & 46.7   & 104.1  \\ \bottomrule
				\end{tabular}
				\label{table:darp312}
				\begin{tablenotes} \scriptsize
					\item[1] F: fragment count; Net: network time, the time required for fragment generation and fragment-based network construction; CPU: CPU time for solver; Time~(s) represents the total time, including preprocessing time, network time, and CPU time for solver. ``No'' indicates that no optimal solution is obtained within the 30-minute time limit. The rows labeled Avg-A and Avg-B represent the average metric values calculated across all instances of type A and type B, respectively.
				\end{tablenotes}
			\end{threeparttable}
		}
	\end{table}

	\begin{table}[H]
		\hspace{0cm}
		\makebox[\textwidth][c]{%
			\begin{threeparttable}
				\caption{Computational performance of PSFF with different fragment sets for MDARP-LPT within 30 minutes (\citeauthor{doi:10.1287/opre.1060.0283}'s instances, $L=4$)}
				\footnotesize
				\setlength{\tabcolsep}{0.2 mm}
				\begin{tabular}{cccccc|cccc|cccc}
					\hline
					&       & \multicolumn{4}{c|}{$S_{RF}$} & \multicolumn{4}{c|}{$S_{FF}$}  & \multicolumn{4}{c}{$S_{MF}$}   \\ \hline
					Name  & OBJ   & F      & Net  & CPU   & Time  & F      & Net   & CPU   & Time  & F      & Net   & CPU   & Time  \\ \hline
					a3-18 & 206.7 & 37     & 8.1  & 0.1   & 8.3   & 33     & 7.2   & 0.1   & 7.4   & 35     & 14.0  & 0.1   & 14.3  \\
					a3-30 & 315.7 & 36     & 8.3  & 0.3   & 8.8   & 36     & 7.1   & 0.3   & 7.6   & 36     & 8.9   & 0.6   & 9.8   \\
					a3-36 & 379.9 & 47     & 7.4  & 0.5   & 10.6  & 44     & 16.9  & 0.6   & 20.2  & 47     & 7.8   & 0.4   & 8.7   \\
					a4-48 & 428.7 & 116    & 10.0 & 1.7   & 13.2  & 109    & 11.7  & 1.8   & 14.8  & 119    & 8.2   & 1.7   & 11.0  \\
					a6-60 & 532.2 & 175    & 10.2 & 5.0   & 17.5  & 156    & 10.4  & 7.4   & 20.0  & 176    & 9.2   & 5.1   & 16.5  \\
					a7-70 & 535.4 & 507    & 16.3 & 19.4  & 40.0  & 505    & 17.2  & 24.3  & 47.0  & 528    & 14.6  & 21.8  & 41.2  \\
					a8-80 & 637.3 & 495    & 21.2 & 37.5  & 65.6  & 586    & 20.8  & 39.5  & 68.9  & 576    & 21.1  & 42.3  & 70.4  \\
					a8-96 & 775.6 & 607    & 16.0 & 87.3  & 116.0 & 551    & 29.8  & 86.5  & 127.3 & 602    & 27.3  & 81.4  & 119.9 \\ \hline
					Avg-A &       & 252.5  &      & 19.0  & 35.0  & 252.5  &       & 20.1  & 39.1  & 264.9  &       & 19.2  & 36.5  \\ \hline
					b3-36 & 364.1 & 146    & 7.5  & 0.5   & 8.6   & 148    & 8.9   & 0.5   & 10.0  & 147    & 9.4   & 0.7   & 12.1  \\
					b4-40 & 345.1 & 370    & 8.4  & 1.4   & 11.0  & 320    & 9.9   & 1.3   & 12.5  & 322    & 18.9  & 1.9   & 22.0  \\
					b5-50 & 417.0 & 1649   & 20.0 & 20.8  & 46.7  & 1682   & 32.2  & 14.7  & 51.4  & 1457   & 30.7  & 18.4  & 53.7  \\
					b6-60 & 517.9 & 2269   & 26.5 & 30.9  & 67.3  & 2084   & 31.6  & 21.6  & 58.8  & 1788   & 35.0  & 23.5  & 65.2  \\
					b7-70 & 512.4 & 1541   & 26.0 & 62.0  & 98.6  & 2049   & 48.9  & 27.6  & 86.4  & 1493   & 53.1  & 48.9  & 110.6 \\
					b8-80 & 601.0 & 2062   & 29.8 & 116.0 & 160.7 & 3133   & 88.4  & 96.0  & 200.7 & 2267   & 93.7  & 101.2 & 209.1 \\
					b8-96 & 714.8 & 6520   & 90.3 & 345.2 & 477.6 & 6582   & 266.8 & 170.8 & 468.5 & 4485   & 275.2 & 304.9 & 611.1 \\ \hline
					Avg-B &       & 2079.6 &      & 82.4  & 124.4 & 2285.4 &       & 47.5  & 126.9 & 1708.4 &       & 71.4  & 154.8 \\ \hline
				\end{tabular}
				\label{table:darp313}
				\begin{tablenotes} \scriptsize
					\item[1] F: fragment count; Net: network time, the time required for fragment generation and fragment-based network construction; CPU: CPU time for solver; Time~(s) represents the total time, including preprocessing time, network time, and CPU time for solver. The rows labeled Avg-A and Avg-B represent the average metric values calculated across all instances of type A and type B, respectively.
				\end{tablenotes}
			\end{threeparttable}
		}
	\end{table}

	\begin{table}[H]
		\hspace{0cm}
		\makebox[\textwidth][c]{%
			\begin{threeparttable}
				\caption{Computational performance of PSFF with different fragment sets for DARP-LPT within 30 minutes (\citeauthor{Riedler2018}'s instances, $L=4$)}
				\footnotesize
				\setlength{\tabcolsep}{0.2mm}
				\begin{tabular}{@{}cccccc|cccc|cccc|cccc|cccc@{}}
					\toprule
					&        & \multicolumn{4}{c|}{$S_{RF}$} & \multicolumn{4}{c|}{$S_{FF}$} & \multicolumn{4}{c|}{$S_{MF}$} & \multicolumn{4}{c|}{$S_{EFF}$} & \multicolumn{4}{c}{$S_{ERF}$}   \\ \midrule
					Name       & OBJ   & F       & Net  & CPU   & Time & F       & Net   & CPU  & Time & F       & Net   & CPU  & Time & F        & Net   & CPU  & Time & F       & Net  & CPU   & Time  \\ \midrule
					30N\_4K\_A & 409.8 & 376     & 0.2  & 0.1   & 0.4  & 372     & 0.4   & 0.1  & 0.5  & 410     & 0.4   & 0.1  & 0.5  & 3499     & 2.5   & 0.1  & 2.9  & 12876   & 3.7  & 0.5   & 5.8   \\
					30N\_4K\_B & 482.4 & 179     & 0.1  & 0.1   & 0.3  & 171     & 0.1   & 0.1  & 0.4  & 180     & 0.2   & 0.2  & 0.4  & 1771     & 1.5   & 0.1  & 1.7  & 6404    & 2.2  & 0.3   & 3.0   \\
					30N\_4K\_C & 436.5 & 603     & 0.4  & 1.2   & 1.9  & 899     & 1.0   & 1.1  & 2.2  & 764     & 1.2   & 2.1  & 3.4  & 5250     & 4.8   & 0.9  & 6.1  & 17664   & 5.1  & 3.1   & 9.6   \\
					30N\_5K\_A & 425.9 & 525     & 0.3  & 0.5   & 1.0  & 759     & 0.7   & 0.4  & 1.4  & 688     & 0.9   & 0.5  & 1.6  & 4823     & 3.5   & 0.5  & 4.3  & 13595   & 4.0  & 0.8   & 6.1   \\
					30N\_5K\_B & 450.9 & 327     & 0.2  & 0.5   & 0.8  & 353     & 0.3   & 0.5  & 0.9  & 365     & 0.5   & 0.5  & 1.1  & 2912     & 2.1   & 0.3  & 2.6  & 8955    & 2.6  & 0.9   & 4.4   \\
					30N\_5K\_C & 460.0 & 466     & 0.3  & 0.7   & 1.1  & 547     & 0.5   & 0.4  & 1.0  & 547     & 0.7   & 0.9  & 1.7  & 3251     & 2.9   & 0.4  & 3.6  & 11851   & 4.0  & 1.5   & 6.5   \\
					40N\_4K\_A & 554.0 & 759     & 0.4  & 1.0   & 1.7  & 965     & 1.4   & 0.5  & 2.1  & 887     & 1.4   & 0.7  & 2.3  & 7629     & 10.3  & 1.5  & 12.7 & 27638   & 8.1  & 3.0   & 13.4  \\
					40N\_4K\_B & 552.7 & 998     & 0.9  & 2.0   & 3.3  & 1375    & 1.9   & 1.3  & 3.5  & 1273    & 1.9   & 1.5  & 3.6  & 14541    & 13.3  & 3.8  & 18.0 & 52456   & 15.5 & 5.3   & 25.6  \\
					40N\_4K\_C & 591.6 & 629     & 0.3  & 0.7   & 1.2  & 657     & 1.3   & 0.7  & 2.2  & 689     & 0.8   & 0.5  & 1.5  & 5536     & 4.3   & 1.1  & 5.9  & 18807   & 5.1  & 1.8   & 9.0   \\
					40N\_5K\_A & 515.0 & 799     & 0.5  & 0.5   & 1.2  & 1276    & 2.0   & 1.5  & 3.8  & 1055    & 1.5   & 1.3  & 3.0  & 8613     & 6.9   & 0.6  & 8.0  & 25871   & 8.1  & 3.1   & 13.5  \\
					40N\_5K\_B & 556.8 & 1023    & 0.6  & 1.4   & 2.3  & 1648    & 2.2   & 0.9  & 3.3  & 1370    & 1.7   & 1.3  & 3.3  & 8184     & 7.2   & 2.2  & 10.0 & 27449   & 8.6  & 8.9   & 19.9  \\
					40N\_5K\_C & 565.7 & 626     & 0.3  & 0.3   & 0.9  & 832     & 1.2   & 0.3  & 1.6  & 760     & 1.1   & 0.2  & 1.5  & 6608     & 5.2   & 1.0  & 6.7  & 22313   & 6.8  & 1.9   & 10.7  \\
					44N\_4K\_A & 580.5 & 1242    & 0.9  & 1.1   & 2.2  & 2015    & 2.6   & 0.8  & 3.6  & 1584    & 2.6   & 0.9  & 3.8  & 16698    & 12.8  & 2.6  & 16.5 & 65445   & 19.4 & 7.0   & 32.1  \\
					44N\_4K\_B & 597.3 & 940     & 0.5  & 0.5   & 1.3  & 1700    & 1.8   & 0.2  & 2.3  & 1312    & 1.9   & 0.7  & 2.8  & 11465    & 9.7   & 1.5  & 12.3 & 32793   & 9.5  & 5.2   & 17.6  \\
					44N\_4K\_C & 588.9 & 1596    & 1.0  & 4.4   & 5.8  & 2653    & 3.2   & 2.0  & 5.4  & 1978    & 3.5   & 2.6  & 6.4  & 15657    & 16.5  & 3.0  & 20.8 & 61526   & 19.3 & 11.9  & 36.7  \\
					44N\_5K\_A & 582.1 & 1162    & 0.7  & 0.5   & 1.6  & 1771    & 2.0   & 0.7  & 2.9  & 1460    & 2.0   & 0.4  & 2.6  & 10167    & 7.5   & 0.8  & 8.9  & 32297   & 10.2 & 1.9   & 14.9  \\
					44N\_5K\_B & 624.9 & 945     & 0.6  & 1.0   & 1.9  & 1426    & 1.8   & 0.7  & 2.7  & 1143    & 1.8   & 1.5  & 3.5  & 10108    & 7.6   & 0.6  & 8.9  & 39795   & 12.8 & 4.0   & 20.2  \\
					44N\_5K\_C & 592.1 & 1056    & 0.6  & 2.0   & 2.8  & 1313    & 2.4   & 1.0  & 3.7  & 1237    & 1.7   & 1.2  & 3.1  & 10907    & 7.5   & 1.8  & 10.1 & 38224   & 11.8 & 5.1   & 20.2  \\
					50N\_4K\_A & 665.0 & 1570    & 1.1  & 2.3   & 3.9  & 2997    & 4.7   & 2.8  & 7.9  & 2273    & 4.7   & 6.9  & 12.1 & 12507    & 11.7  & 3.8  & 16.5 & 41095   & 13.4 & 15.7  & 34.6  \\
					50N\_4K\_B & 685.3 & 1322    & 0.7  & 0.7   & 1.9  & 2094    & 3.2   & 0.4  & 3.9  & 1732    & 3.3   & 0.4  & 4.0  & 18800    & 13.8  & 0.8  & 15.9 & 59798   & 29.0 & 6.0   & 41.5  \\
					50N\_4K\_C & 641.7 & 1146    & 0.8  & 2.2   & 3.4  & 1888    & 2.8   & 1.3  & 4.3  & 1525    & 2.8   & 1.3  & 4.5  & 14709    & 10.8  & 2.2  & 14.0 & 46530   & 18.4 & 17.5  & 41.0  \\
					50N\_5K\_A & 645.5 & 2245    & 2.3  & 10.2  & 13.2 & 5473    & 7.2   & 4.5  & 12.1 & 3267    & 9.1   & 5.3  & 14.9 & 32723    & 26.7  & 5.8  & 34.5 & 124110  & 52.0 & 87.2  & 152.0 \\
					50N\_5K\_B & 666.2 & 1393    & 0.8  & 3.6   & 5.0  & 2623    & 4.5   & 2.1  & 7.0  & 1896    & 3.9   & 1.2  & 5.5  & 21982    & 16.3  & 3.3  & 21.5 & 71971   & 25.8 & 12.9  & 46.2  \\
					50N\_5K\_C & 667.8 & 1163    & 0.8  & 4.7   & 5.9  & 1438    & 3.1   & 6.8  & 10.3 & 1347    & 2.6   & 4.4  & 7.4  & 18851    & 13.5  & 4.6  & 19.5 & 77905   & 24.8 & 38.4  & 72.9  \\
					60N\_4K\_A & 814.5 & 1830    & 1.2  & 7.2   & 8.9  & 3313    & 4.8   & 4.5  & 9.7  & 2514    & 4.8   & 4.7  & 9.9  & 30150    & 23.0  & 4.0  & 29.3 & 88197   & 34.2 & 23.1  & 66.1  \\
					60N\_4K\_B & 727.3 & 4533    & 4.5  & 21.2  & 26.9 & 10314   & 18.3  & 16.3 & 35.5 & 5961    & 18.0  & 15.0 & 33.7 & 68432    & 58.2  & 10.5 & 72.8 & 257386  & 86.3 & 142.3 & 252.1 \\
					60N\_4K\_C & 747.1 & 2588    & 1.7  & 16.2  & 18.7 & 5205    & 12.6  & 4.8  & 18.2 & 3409    & 9.2   & 5.8  & 15.6 & 40287    & 32.3  & 6.2  & 41.3 & 141157  & 43.3 & 58.6  & 117.5 \\
					60N\_5K\_A & 743.2 & 2618    & 1.7  & 6.4   & 9.1  & 5543    & 12.9  & 0.9  & 14.5 & 3726    & 8.8   & 2.1  & 11.4 & 44007    & 35.9  & 4.9  & 43.8 & 146211  & 45.4 & 33.5  & 96.0  \\
					60N\_5K\_B & 800.8 & 2260    & 1.7  & 2.8   & 5.4  & 4585    & 9.2   & 3.6  & 13.4 & 3103    & 8.1   & 1.8  & 10.5 & 39063    & 30.2  & 3.8  & 37.0 & 126061  & 41.1 & 62.5  & 115.6 \\
					60N\_5K\_C & 822.7 & 2125    & 2.0  & 6.4   & 9.1  & 3647    & 6.7   & 1.7  & 8.9  & 2633    & 6.6   & 2.8  & 9.8  & 34999    & 27.5  & 5.3  & 35.4 & 128702  & 42.4 & 21.8  & 75.8  \\ \midrule
					Avg        &       & 1301.5  &      & 3.4   & 4.8  & 2328.4  &       & 2.1  & 6.3  & 1702.9  &       & 2.3  & 6.2  & 17471.0  &       & 2.6  & 18.0 & 60836.1 &      & 19.5  & 46.0  \\ \bottomrule
				\end{tabular}
				\label{table:darp321}
				\begin{tablenotes} \scriptsize
					\item[1] F: fragment count; Net: network time, the time required for fragment generation and fragment-based network construction; CPU: CPU time for solver; Time~(s) represents the total time, including preprocessing time, network time, and CPU time for solver. Avg represents the average metric values calculated across all instances.
				\end{tablenotes}
			\end{threeparttable}
		}
	\end{table}

	\begin{table}[H]
		\hspace{0cm}
		\makebox[\textwidth][c]{%
			\begin{threeparttable}
				\caption{Computational performance of PSFF with different fragment sets for DARP-LPT within 30 minutes (\citeauthor{Riedler2018}'s instances, $L=6$)}
				\footnotesize
				\setlength{\tabcolsep}{0.2mm}
				\begin{tabular}{@{}cccccc|cccc|cccc|cccc|cccc@{}}
					\toprule
					&        & \multicolumn{4}{c|}{$S_{RF}$} & \multicolumn{4}{c|}{$S_{FF}$} & \multicolumn{4}{c|}{$S_{MF}$} & \multicolumn{4}{c|}{$S_{EFF}$} & \multicolumn{4}{c}{$S_{ERF}$}   \\ \midrule
					Name       & OBJ   & F       & Net & CPU   & Time  & F       & Net   & CPU   & Time  & F      & Net   & CPU   & Time  & F       & Net   & CPU   & Time   & F    & Net    & CPU    & Time  \\ \midrule
					30N\_4K\_A & 392.2 & 376     & 0.2 & 0.5   & 1.3   & 386     & 1.2   & 1.1   & 2.4   & 411    & 0.7   & 0.3   & 1.2   & 4188    & 3.6   & 0.8   & 5.0    & 12876  & 5.1    & 1.3    & 12.0 \\
					30N\_4K\_B & 459.7 & 179     & 0.1 & 0.1   & 0.3   & 174     & 0.6   & 0.4   & 1.4   & 180    & 0.2   & 0.2   & 0.4   & 1775    & 1.4   & 0.5   & 2.1    & 6404   & 1.8    & 0.4    & 3.1    \\
					30N\_4K\_C & 419.5 & 603     & 0.3 & 20.9  & 21.8  & 2565    & 11.4  & 15.8  & 27.5  & 777    & 5.4   & 4.9   & 10.4  & 13862   & 18.4  & 11.2  & 31.3   & 17664  & 8.3    & 58.8   & 69.8   \\
					30N\_5K\_A & 408.5 & 525     & 0.3 & 4.3   & 4.9   & 1396    & 2.7   & 10.9  & 13.8  & 694    & 2.3   & 5.9   & 8.5   & 9187    & 9.1   & 2.5   & 12.9   & 13595  & 4.8    & 17.0   & 24.8   \\
					30N\_5K\_B & 425.2 & 327     & 0.3 & 1.2   & 1.7   & 499     & 1.0   & 1.4   & 2.6   & 366    & 1.1   & 3.6   & 5.0   & 4269    & 7.5   & 1.9   & 10.8   & 8955   & 2.6    & 4.4    & 9.3    \\
					30N\_5K\_C & 445.1 & 466     & 0.3 & 1.7   & 2.3   & 857     & 1.6   & 1.7   & 3.5   & 564    & 2.2   & 1.7   & 4.1   & 5210    & 9.4   & 1.8   & 12.3   & 11851  & 4.5    & 11.2   & 17.5   \\
					40N\_4K\_A & 525.3 & 759     & 0.5 & 3.8   & 4.8   & 2277    & 7.0   & 3.7   & 11.1  & 918    & 5.8   & 2.4   & 8.5   & 15364   & 25.8  & 7.9   & 36.0   & 27638  & 8.8    & 80.9   & 95.1   \\
					40N\_4K\_B & 522.2 & 998     & 0.7 & 3.8   & 5.0   & 1752    & 8.7   & 3.1   & 12.2  & 1288   & 6.8   & 3.3   & 10.5  & 25987   & 29.0  & 10.3  & 42.4   & 52456  & 18.0   & 26.4   & 52.7   \\
					40N\_4K\_C & 557.5 & 629     & 0.5 & 7.5   & 8.6   & 822     & 2.1   & 4.2   & 6.5   & 696    & 1.9   & 9.0   & 11.2  & 6472    & 7.2   & 1.6   & 10.1   & 18807  & 6.1    & 19.9   & 30.1   \\
					40N\_5K\_A & 477.5 & 799     & 0.5 & 1.6   & 2.7   & 3252    & 10.0  & 1.8   & 12.3  & 1069   & 12.4  & 2.3   & 15.0  & 20806   & 27.0  & 2.2   & 31.8   & 25871  & 12.7   & 19.6   & 38.1   \\
					40N\_5K\_B & 528.9 & 1023    & 0.7 & 12.0  & 13.2  & 3705    & 7.6   & 23.3  & 31.5  & 1397   & 7.7   & 17.6  & 25.7  & 16061   & 19.9  & 13.6  & 35.8   & 27449  & 8.9    & 79.8   & 94.7   \\
					40N\_5K\_C & 536.8 & 626     & 0.3 & 0.4   & 1.1   & 1768    & 4.0   & 0.3   & 4.7   & 769    & 3.9   & 0.4   & 4.6   & 12393   & 14.8  & 0.8   & 17.4   & 22313  & 6.7    & 10.6   & 21.6   \\
					44N\_4K\_A & 541.6 & 1242    & 0.9 & 4.5   & 5.9   & 6294    & 14.4  & 4.2   & 19.4  & 1623   & 13.8  & 4.3   & 18.5  & 50962   & 54.8  & 10.6  & 70.6   & 65445  & 29.0   & 185.5  & 225.2  \\
					44N\_4K\_B & 562.4 & 940     & 0.6 & 8.1   & 9.4   & 6362    & 9.6   & 10.2  & 20.5  & 1341   & 13.2  & 11.1  & 24.7  & 32137   & 33.7  & 6.1   & 43.0   & 32793  & 12.3   & 155.9  & 174.3  \\
					44N\_4K\_C & 551.6 & 1596    & 1.1 & 32.2  & 33.9  & 12758   & 29.5  & 19.6  & 50.3  & 2125   & 33.3  & 25.5  & 59.4  & 61590   & 78.3  & 43.1  & 127.7  & 61526  & 23.0   & 247.0  & 279.7  \\
					44N\_5K\_A & 554.9 & 1162    & 0.7 & 13.3  & 14.7  & 4183    & 7.2   & 15.7  & 23.4  & 1491   & 8.6   & 8.6   & 17.7  & 23984   & 25.7  & 5.0   & 33.4   & 32297  & 13.0   & 142.7  & 161.3  \\
					44N\_5K\_B & 589.3 & 945     & 0.6 & 10.7  & 11.9  & 3677    & 8.7   & 9.8   & 19.3  & 1167   & 8.4   & 9.9   & 18.7  & 22135   & 28.0  & 4.4   & 34.7   & 39795  & 15.4   & 19.2   & 40.8   \\
					44N\_5K\_C & 568.4 & 1056    & 0.7 & 26.6  & 27.9  & 2218    & 6.6   & 22.7  & 29.8  & 1277   & 6.3   & 24.7  & 31.4  & 23095   & 25.7  & 16.4  & 44.7   & 38224  & 12.3   & 105.6  & 125.6  \\
					50N\_4K\_A & 628.4 & 1570    & 1.3 & 31.1  & 33.3  & 7105    & 18.3  & 45.2  & 64.2  & 2310   & 16.3  & 28.3  & 45.1  & 24443   & 39.2  & 12.1  & 54.2   & 41095  & 15.3   & 355.3  & 378.9  \\
					50N\_4K\_B & 639.0 & 1322    & 0.8 & 6.0   & 7.6   & 7106    & 20.3  & 5.0   & 26.2  & 1755   & 18.2  & 4.0   & 22.7  & 51665   & 53.1  & 7.5   & 65.8   & 59798  & 21.5   & 172.1  & 203.3  \\
					50N\_4K\_C & 606.8 & 1146    & 0.7 & 13.7  & 15.1  & 5169    & 13.5  & 15.3  & 29.6  & 1530   & 11.3  & 14.6  & 26.4  & 30064   & 32.4  & 15.7  & 51.6   & 46530  & 15.3   & 500.0  & 524.0  \\
					50N\_5K\_A & 601.8 & 2245    & 1.7 & 177.2 & 179.9 & 45134   & 120.3 & 294.2 & 418.0 & 3506   & 108.9 & 127.1 & 236.8 & 221956  & 322.3 & 211.4 & 553.3  & 124110 & 53.7   & 1723.0 & 1800.9 \\
					50N\_5K\_B & 624.4 & 1393    & 1.0 & 55.3  & 57.0  & 11381   & 29.3  & 32.0  & 62.9  & 1935   & 29.0  & 26.7  & 56.2  & 81754   & 99.2  & 102.5 & 212.1  & 71971  & 26.2   & 399.7  & 439.6  \\
					50N\_5K\_C & 619.7 & 1163    & 0.7 & 46.2  & 47.6  & 2141    & 11.6  & 20.5  & 32.5  & 1362   & 11.1  & 22.4  & 34.4  & 33933   & 35.0  & 32.6  & 71.1   & 77905  & 32.9   & 289.3  & 338.5  \\
					60N\_4K\_A & 762.8 & 1830    & 1.6 & 16.3  & 19.7  & 10331   & 35.1  & 15.1  & 51.3  & 2592   & 33.0  & 5.9   & 39.6  & 93547   & 133.0 & 57.8  & 200.7  & 88197  & 29.0   & 315.9  & 364.2  \\
					60N\_4K\_B & 680.1 & 4533    & 5.9 & 205.6 & 214.8 & 103503  & 273.4 & 337.3 & 619.1 & 6569   & 273.1 & 113.5 & 388.1 & 553669  & 797.0 & 586.0 & 1432.7 & \multicolumn{4}{c}{No}         \\
					60N\_4K\_C & 698.1 & 2588    & 2.1 & 270.2 & 273.7 & 32289   & 91.7  & 234.5 & 329.3 & 3642   & 88.7  & 198.2 & 288.0 & 212470  & 303.8 & 147.1 & 472.8  & \multicolumn{4}{c}{No}         \\
					60N\_5K\_A & 698.8 & 2618    & 2.3 & 37.1  & 41.2  & 36234   & 107.6 & 78.8  & 189.2 & 4011   & 105.0 & 27.2  & 133.1 & 229204  & 327.9 & 244.7 & 594.6  & \multicolumn{4}{c}{No}         \\
					60N\_5K\_B & 748.1 & 2260    & 1.7 & 133.2 & 136.6 & 25308   & 65.7  & 152.4 & 220.3 & 3234   & 63.8  & 128.8 & 193.4 & 166511  & 245.4 & 203.4 & 467.3  & \multicolumn{4}{c}{No}         \\
					60N\_5K\_C & 768.6 & 2125    & 1.6 & 167.4 & 170.7 & 14714   & 51.2  & 107.4 & 160.1 & 2878   & 52.3  & 96.6  & 149.6 & 151402  & 222.1 & 169.3 & 407.0  & 128702 & 44.6   & 505.3  & 572.4  \\ \midrule
					Avg        &       & 1301.5  &     & 43.8  & 45.6  & 11845.3 &       & 49.6  & 83.2  & 1782.6 &       & 31.0  & 63.0  & 73336.5 &       & 64.4  & 172.8  & \multicolumn{4}{c}{26/30 solved}         \\ \bottomrule
				\end{tabular}\label{table:darp322}
				\begin{tablenotes} \scriptsize
					\item[1] F: fragment count; Net: network time, the time required for fragment generation and fragment-based network construction; CPU: CPU time for solver; Time~(s) represents the total time, including preprocessing time, network time, and CPU time for solver. Avg represents the average metric values calculated across all solvable instances. ``No'' indicates that no optimal solution is obtained within the 30-minute time limit.
				\end{tablenotes}
			\end{threeparttable}
		}
	\end{table}

	\begin{table}[H]
		\hspace{0cm}
		\makebox[\textwidth][c]{%
			\begin{threeparttable}
				\caption{Computational performance of PSFF with different fragment sets for MDARP-LPT within 30 minutes (\citeauthor{Riedler2018}'s instances, $L=4$)}
				\footnotesize
				\setlength{\tabcolsep}{0.2mm}
				\begin{tabular}{@{}cccccc|cccc|cccc@{}}
					\toprule
					&       & \multicolumn{4}{c|}{$S_{RF}$} & \multicolumn{4}{c|}{$S_{FF}$} & \multicolumn{4}{c}{$S_{MF}$} \\ \midrule
					Name       & OBJ   & F      & Net  & CPU   & Time  & F      & Net  & CPU   & Time  & F      & Net  & CPU  & Time  \\ \midrule
					30N\_4K\_A & 267.1 & 146    & 10.7 & 1.4   & 13.3  & 134    & 8.1  & 0.9   & 9.5   & 156    & 8.1  & 0.7  & 9.2   \\
					30N\_4K\_B & 313.1 & 72     & 8.9  & 1.0   & 10.4  & 66     & 7.2  & 0.4   & 7.8   & 72     & 7.2  & 0.4  & 9.9   \\
					30N\_4K\_C & 280.4 & 146    & 9.0  & 3.3   & 12.8  & 132    & 7.3  & 4.0   & 12.1  & 149    & 7.0  & 2.1  & 9.4   \\
					30N\_5K\_A & 251.1 & 194    & 11.8 & 2.2   & 15.7  & 205    & 12.4 & 1.2   & 14.1  & 219    & 12.0 & 1.6  & 14.1  \\
					30N\_5K\_B & 295.5 & 150    & 10.4 & 1.3   & 12.2  & 140    & 8.2  & 0.5   & 9.2   & 155    & 7.2  & 0.8  & 8.5   \\
					30N\_5K\_C & 299.4 & 180    & 8.4  & 1.2   & 10.3  & 195    & 7.9  & 1.6   & 9.9   & 206    & 7.2  & 4.8  & 12.4  \\
					40N\_4K\_A & 371.2 & 259    & 12.0 & 2.5   & 16.0  & 266    & 7.6  & 1.7   & 10.1  & 273    & 11.0 & 1.5  & 13.6  \\
					40N\_4K\_B & 311.3 & 441    & 10.6 & 9.1   & 21.7  & 445    & 12.0 & 2.7   & 15.7  & 479    & 17.8 & 5.2  & 24.1  \\
					40N\_4K\_C & 374.1 & 197    & 10.6 & 0.9   & 12.8  & 182    & 14.3 & 0.7   & 15.7  & 192    & 8.5  & 0.9  & 10.2  \\
					40N\_5K\_A & 294.6 & 271    & 9.1  & 2.7   & 13.1  & 327    & 10.0 & 2.0   & 13.1  & 321    & 8.3  & 1.5  & 10.8  \\
					40N\_5K\_B & 340.8 & 327    & 11.6 & 2.6   & 15.8  & 395    & 16.8 & 2.2   & 19.9  & 373    & 7.9  & 1.9  & 10.7  \\
					40N\_5K\_C & 366.7 & 229    & 9.3  & 6.1   & 16.6  & 226    & 9.5  & 4.4   & 14.8  & 239    & 7.6  & 2.5  & 10.9  \\
					44N\_4K\_A & 369.2 & 361    & 13.7 & 4.8   & 20.4  & 437    & 9.6  & 3.5   & 14.3  & 423    & 11.3 & 4.9  & 17.5  \\
					44N\_4K\_B & 381.4 & 383    & 10.2 & 10.6  & 23.0  & 547    & 13.2 & 3.1   & 17.5  & 498    & 15.0 & 7.0  & 23.4  \\
					44N\_4K\_C & 390.5 & 503    & 11.5 & 20.3  & 34.0  & 668    & 11.6 & 11.5  & 24.6  & 601    & 12.9 & 21.3 & 36.4  \\
					44N\_5K\_A & 340.6 & 533    & 13.2 & 4.9   & 19.9  & 719    & 13.1 & 1.7   & 16.3  & 614    & 12.8 & 2.7  & 17.5  \\
					44N\_5K\_B & 368.0 & 326    & 9.1  & 6.9   & 17.4  & 378    & 10.5 & 6.2   & 18.1  & 375    & 10.8 & 10.4 & 22.9  \\
					44N\_5K\_C & 369.3 & 334    & 14.6 & 9.7   & 25.6  & 319    & 16.0 & 7.3   & 24.5  & 352    & 11.8 & 8.1  & 21.6  \\
					50N\_4K\_A & 396.8 & 475    & 10.3 & 13.2  & 26.0  & 535    & 13.5 & 6.3   & 21.6  & 542    & 23.0 & 10.9 & 36.2  \\
					50N\_4K\_B & 423.9 & 539    & 12.8 & 11.7  & 27.5  & 686    & 17.9 & 6.9   & 26.8  & 632    & 21.8 & 13.3 & 37.6  \\
					50N\_4K\_C & 386.5 & 519    & 12.7 & 9.2   & 24.3  & 802    & 13.5 & 7.9   & 23.7  & 681    & 19.0 & 15.4 & 37.5  \\
					50N\_5K\_A & 388.7 & 962    & 18.7 & 28.1  & 50.6  & 1928   & 22.7 & 36.7  & 64.4  & 1283   & 24.2 & 31.8 & 59.9  \\
					50N\_5K\_B & 407.2 & 612    & 17.5 & 25.3  & 45.7  & 965    & 17.4 & 28.7  & 49.3  & 769    & 21.5 & 48.8 & 73.2  \\
					50N\_5K\_C & 401.2 & 415    & 16.1 & 16.2  & 34.2  & 458    & 12.8 & 6.1   & 21.0  & 473    & 13.1 & 9.6  & 24.7  \\
					60N\_4K\_A & 509.6 & 622    & 11.2 & 12.5  & 27.4  & 741    & 18.1 & 7.6   & 28.9  & 728    & 22.2 & 8.8  & 34.7  \\
					60N\_4K\_B & 437.0 & 1502   & 18.3 & 88.7  & 114.9 & 2657   & 38.9 & 58.1  & 104.5 & 1879   & 35.1 & 26.8 & 68.5  \\
					60N\_4K\_C & 475.4 & 891    & 12.9 & 116.9 & 135.6 & 1432   & 28.1 & 161.1 & 194.2 & 1162   & 22.8 & 40.1 & 67.8  \\
					60N\_5K\_A & 476.8 & 736    & 10.1 & 64.9  & 80.7  & 1203   & 21.9 & 32.5  & 58.4  & 985    & 27.1 & 92.6 & 124.5 \\
					60N\_5K\_B & 469.5 & 781    & 12.5 & 25.6  & 43.0  & 1194   & 42.8 & 15.7  & 63.3  & 987    & 29.1 & 34.6 & 70.0  \\
					60N\_5K\_C & 531.9 & 535    & 10.3 & 12.0  & 26.3  & 634    & 16.4 & 14.2  & 33.9  & 593    & 20.4 & 18.2 & 41.8  \\ \midrule
					Avg        &       & 454.7  &      & 17.2  & 31.6  & 633.9  &      & 14.6  & 31.9  & 547.0  &      & 14.3 & 32.0  \\ \bottomrule
				\end{tabular}
				\label{table:darp323}
				\begin{tablenotes} \scriptsize
					\item[1] F: fragment count; Net: network time, the time required for fragment generation and fragment-based network construction; CPU: CPU time for solver; Time~(s) represents the total time, including preprocessing time, network time, and CPU time for solver. Avg represents the average metric values calculated across all instances.
				\end{tablenotes}
			\end{threeparttable}
		}
	\end{table}

	\section{Full computational results of FFF for DARP-LPT }\label{resultDARPLPTFFF}
	
	\setcounter{table}{0}
	
	Table~\ref{table:darp11} describe the computational performance of FFF with different fragment sets for DARP-LPT within 30 minutes.
	
	\begin{table}[H]
		\hspace{0cm}
		\makebox[\textwidth][c]{%
			\begin{threeparttable}
				\caption{Computational performance of FFF with different fragment sets for DARP-LPT within 30 minutes (\citeauthor{doi:10.1287/opre.1060.0283}'s instances, $L=4$)}
				\footnotesize
				\setlength{\tabcolsep}{0.5mm}
				\begin{tabular}{@{}cccccccc|ccccccc@{}}
					\toprule
					\multicolumn{1}{l}{} & \multicolumn{7}{c|}{$S_{RF}$}                            & \multicolumn{7}{c}{$S_{FF}$}                              \\ \midrule
					Name                 & F     & Net & CPU    & Time   & OBJ    & LB     & Gap    & F     & Net  & CPU    & Time   & OBJ    & LB     & Gap    \\\midrule
					a3-18                & 73    & 0.3 & 0.0    & 0.4    & 302.5  & 302.5  & 0.00\% & 64    & 0.1  & 0.0    & 0.2    & 302.5  & 302.5  & 0.00\% \\
					a3-30                & 93    & 0.1 & 1799.9 & 1800.0 & 494.8  & 491.0  & 0.75\% & 89    & 0.1  & 1799.9 & 1800.0 & 494.8  & 491.1  & 0.74\% \\
					a3-36                & 121   & 0.1 & 1800.5 & 1800.0 & 574.4  & 560.6  & 2.40\% & 112   & 0.2  & 1797.6 & 1800.0 & 574.4  & 565.0  & 1.63\% \\
					a4-48                & 318   & 0.2 & 1798.9 & 1800.0 & 697.9  & 686.1  & 1.69\% & 312   & 0.5  & 1799.4 & 1800.0 & 697.9  & 687.4  & 1.50\% \\
					a6-60                & 601   & 0.4 & 1793.4 & 1800.0 & 869.0  & 843.2  & 2.97\% & 648   & 1.4  & 1795.5 & 1800.0 & 864.9  & 843.4  & 2.48\% \\
					a7-70                & 1036  & 0.8 & 1796.3 & 1800.0 & 986.3  & 958.5  & 2.82\% & 1273  & 2.6  & 1792.3 & 1800.0 & 986.6  & 956.5  & 3.05\% \\
					a8-80                & 1088  & 0.8 & 1789.1 & 1800.0 & 1064.6 & 1016.4 & 4.53\% & 1473  & 5.1  & 1785.7 & 1800.0 & 1071.1 & 1019.3 & 4.84\% \\
					a8-96                & 1465  & 1.0 & 1756.7 & 1800.0 & 1354.8 & 1296.8 & 4.28\% & 1684  & 7.3  & 1759.3 & 1800.0 & 1353.6 & 1302.7 & 3.76\% \\ \midrule
					Avg-A                & 599.4 &     & 1566.9 & 1575.1 &        &        & 2.43\% & 706.9 &      & 1566.2 & 1575.0 &        &        & 2.25\% \\ \midrule
					b3-36                & 169   & 0.1 & 1799.9 & 1800.0 & 576.5  & 566.2  & 1.79\% & 161   & 0.1  & 1799.7 & 1800.0 & 576.5  & 567.7  & 1.53\% \\
					b4-40                & 311   & 0.2 & 1799.5 & 1800.0 & 635.2  & 630.4  & 0.76\% & 203   & 0.4  & 1799.1 & 1800.0 & 635.2  & 631.1  & 0.64\% \\
					b5-50                & 518   & 0.4 & 1798.6 & 1800.0 & 789.1  & 766.2  & 2.91\% & 803   & 1.2  & 1797.6 & 1800.0 & 789.1  & 768.3  & 2.63\% \\
					b6-60                & 940   & 0.6 & 1793.3 & 1800.0 & 931.4  & 895.5  & 3.86\% & 1218  & 2.6  & 1792.3 & 1800.0 & 931.7  & 900.7  & 3.33\% \\
					b7-70                & 960   & 0.7 & 1796.4 & 1800.0 & 948.0  & 921.6  & 2.79\% & 1655  & 3.2  & 1789.5 & 1800.0 & 950.4  & 920.7  & 3.13\% \\
					b8-80                & 1195  & 2.0 & 1790.3 & 1800.0 & 1105.8 & 1071.8 & 3.08\% & 1813  & 5.1  & 1788.6 & 1800.0 & 1104.8 & 1072.0 & 2.97\% \\
					b8-96                & \multicolumn{7}{c|}{No}                                  & 6540  & 33.2 & 1734.9 & 1800.0 & 1322.6 & 1260.9 & 4.67\% \\ \midrule
					Avg-B                & 682.2 &     & 1796.3 & 1800.0 &        &        & 2.53\% & 975.5 &      & 1794.5 & 1800.0 &        &        & 2.37\% \\ \bottomrule
				\end{tabular}
				\label{table:darp11}
				\begin{tablenotes} \scriptsize
					\item[1] F: fragment count; Net: network time, the time required for fragment generation and fragment-based network construction; MIP: CPU time for solver; Time~(s) represents the total time, including preprocessing time, network time, and CPU time for solver. The rows labeled Avg-A and Avg-B represent the average metric values calculated across all instances of type A and type B, respectively.  Avg-B does not involve the information of ``b8-96'' for both $S_{RF}$ and $S_{FF}$ since the former cannot lead to any feasible solution within the 30 minutes.
				\end{tablenotes}
			\end{threeparttable}
		}
	\end{table}

\end{document}